\documentclass[pra,twocolumn,longbibliography]{revtex4}

\usepackage[utf8]{inputenc}
\usepackage{bm}
\usepackage{mathtools}
\usepackage{bbold}
\usepackage{amsmath}
\usepackage{amssymb}
\usepackage{hyperref}
\usepackage[normalem]{ulem}
\usepackage{color}
\usepackage[dvipsnames]{xcolor}

\begin{document}

\title{Rotor/spin-wave theory for quantum spin models with U(1) symmetry}
\author{Tommaso Roscilde, Tommaso Comparin, and Fabio Mezzacapo}
\affiliation{Univ Lyon, Ens de Lyon, CNRS, Laboratoire de Physique, F-69342 Lyon, France}


\begin{abstract}
The static and dynamic properties of finite-size lattice quantum spin models which spontaneously break a continuous $U(1)$ symmetry in the thermodynamic limit are of central importance for a wide variety of physical systems, from condensed matter to quantum simulation. Such systems are characterized by a Goldstone excitation branch, terminating in a zero mode whose theoretical treatment within a linearized approach leads to divergencies on finite-size systems, revealing that the assumption of symmetry breaking is ill-defined away from the thermodynamic limit. In this work we show that, once all its non-linearities are taken into account, the zero mode corresponds exactly to a U(1) quantum rotor, related to the Anderson tower of states expected in systems showing symmetry breaking in the thermodynamic limit. The finite-momentum modes, when weakly populated, can be instead safely linearized (namely treated within spin-wave theory) and effectively decoupled from the zero mode. This picture leads to an approximate separation of variables between rotor and spin-wave ones, which allows for a correct description of the ground-state and low-energy physics. Most importantly, it offers a quantitative treatment of the finite-size non-equilibrium dynamics -- following a quantum quench -- dominated by the zero mode, for which a linearized approach fails after a short time. Focusing on the 2$d$ XX model with power-law decaying interactions, we compare our equilibrium predictions with unbiased quantum Monte Carlo results and exact diagonalization; and our non-equilibrium results with time-dependent variational Monte Carlo. The agreement is remarkable for all interaction ranges, and it improves the longer the range. Our rotor/spin-wave theory defines a successful strategy for the application of spin-wave theory and its extensions to finite-size systems at equilibrium or away from it. 
\end{abstract}
\maketitle

\section{Introduction}
\label{s.intro}

 Lattice quantum spin models \cite{Auerbachbook,Blundellbook} occupy a central spot in the field of many-body quantum physics, offering a quantitative description of magnetism in insulating materials; as well as a set  of paradigmatic models for quantum statistical mechanics. More recently, the study of quantum spin models has entered a new
  dimension in the field of quantum simulation \cite{Georgescuetal2014}, as such models can be implemented in systems of interacting qubits ($S=1/2$ spins) or qudits ($S>1/2$ spins), realized by ultracold atoms in optical lattices \cite{Mazurenko2017,Lepoutreetal2019,Jepsen2020,Chomazetal2022,Christakisetal2023}, trapped ions \cite{Monroe2021RMP}, arrays of Rydberg atoms \cite{BrowaeysL2020}, or superconducting circuits \cite{Juanjobook}, to cite a few relevant platforms.
      
 Quantum simulation introduces two new fundamental aspects to the study of quantum magnetism: 1) quantum simulators realize mesoscopic spin assemblies (with particle numbers $N$  ranging from $\sim 10$ to $\sim 10^4$, depending on the platform), whose finite-size nature is a fundamental feature, and not necessarily a limitation; and 2) quantum simulators naturally realize unitary non-equilibrium dynamics of quantum spin models, over times that are sufficiently long for the finiteness of the system size to play a role in the dynamics. The faithful  theoretical study of non-equilibrium quantum dynamics requires the ability to describe the evolution of correlations and entanglement; and in particular to do that accounting for non-linearities, which are essential in finite-size dynamics -- as we will further elaborate below. 
In the context of quantum spin models, the simplest approach to deal with quantum correlations and entanglement beyond the mean-field level is linear spin-wave (LSW) theory \cite{Auerbachbook,Levybook}, which approximately maps the quantum spin problem onto a quadratic bosonic Hamiltonian describing linearized quantum fluctuations around a classically ordered state. LSW theory rests on the assumption of spontaneous symmetry breaking (SSB), namely the picture by which the dynamics of a physical system remains confined in the vicinity of the classically ordered configuration, developing weak oscillations around the ordered state. This picture, allowing for the linearization of the dynamics, is fully justified in the thermodynamic limit, in which SSB is properly realized. On the other hand the image of linear fluctuations around an ordered state can substantially fail in finite-size systems, implying the disruption of classical order and the appearance of fully non-linear quantum effects; such effects result typically in richer forms of entanglement than those allowed for by linear quantum fluctuations. 

The dynamical disruption of long-range order in finite-size systems is particularly serious in the case of continuous symmetries -- and in this work we will focus on translationally invariant systems with U(1) symmetry, namely on uniform \emph{planar magnets}, in which two spin components are equally coupled, and this coupling dominates the energetics of the system at low energies. Systems breaking a U(1) symmetry possess a gapless Goldstone branch of excitations (correctly accounted for by LSW theory), which terminates in a \emph{zero mode} -- namely a zero frequency mode at zero wave-vector, associated with the dynamical restoration of the U(1) symmetry. Yet the failure of conventional LSW theory on a finite-size system is signaled by the significant difficulties encountered in the treatment of this zero mode. First and foremost, a fully gapless spectrum is not allowed on a finite-size system in the absence of accidental degeneracies. In fact the low-energy spectrum of a system breaking a continuous symmetry is well known to feature instead an Anderson tower of states (ToS), namely a discrete spectrum of non-linear excitations akin to that of a quantum rotor \cite{Anderson1997,Lauchli2016,Tasaki2018JSP}. Ignoring this aspect, and including naively the gapless zero mode in the system, leads to divergencies in the momentum-space sums that determine some of the most basic predictions of LSW theory, above all the magnitude of the order parameter. 

In the face of this problem, three strategies can be contemplated in order to formulate a finite-size LSW theory: 1) \emph{zero-mode removal}:  the zero mode can be simply eliminated from the treatment, on the account that in the thermodynamic limit its contribution to the momentum-space integrals  would vanish. This approach is justified as a way to mimic the thermodynamic limit using a finite-size system, but it fails to capture the specific aspects brought about by a finite size;  2) \emph{gapping out the zero mode}: the zero mode can be still included in the LSW description, but its pathological aspects are cured by the application of a field coupling to the order parameter, which gaps the mode out \cite{Songetal2011, FrerotR2015,Frerot2017PRB}.  This approach is also at the basis of non-linear extensions of LSW theory (namely the so-called modified spin-wave theory \cite{Takahashi1989}); 
3) \emph{separate treatment of the zero mode}: in translationally invariant systems the zero mode can be formally separated from the rest of the modes within LSW theory, as it stems from bosonic operators creating and destroying zero-momentum bosons, which to quadratic order are decoupled from finite-momentum operators because of momentum conservation. The zero-momentum bosons can be treated differently from the finite-momentum ones, and they can be cast in terms of bosonic quadratures instead of being Bogolyubov diagonalized; this approach has been put forward in Refs.~\cite{ZhongS1993,Trumperetal2000,Capriotti2003}, and successfully applied to SU(2)-symmetric Heisenberg antiferromagnets. 

In this work we adopt the third strategy of a separate treatment of the zero mode, namely of the zero-momentum operators; and we push this approach far beyond the picture of a quadratic bosonic Hamiltonian, so as to account for the quantum non-linearities associated with the disruption of classical order in a finite-size system. Here is a summary of our main results: 
\begin{itemize}
\item we show that all the (linear and non-linear) terms in the bosonic Hamiltonian involving exclusively zero-momentum bosons reconstruct the Hamiltonian of a \emph{U(1) quantum-rotor variable}, namely a giant spin of length $NS$, with moment of inertia $\sim N$. The quantum-rotor Hamiltonian exhibits the energy spectrum of the Anderson ToS, expected in a finite-size system. The equilibrium low-energy configurations of finite-size systems, as well as the non-equilibrium ones reached during quench dynamics, imply a depolarization of the rotor, corresponding to a \emph{macroscopic} population of the zero-momentum bosons \footnote{We would not call this phenomenon a \emph{condensation} of zero-momentum bosons, as the bosons in question are quasi-particles whose number is not conserved, and which can dynamically go from zero to a macroscopic value. In the latter situation long-range order is disrupted in the system; hence the picture of a condensate, which is associated instead with long-range phase coherence, may therefore be confusing.}, and therefore the treatment of its non-linearities is essential. 
\item on the other hand the finite-momentum bosonic modes can be assumed to remain only weakly populated (${\cal O}(1)$ populations); therefore the Hamiltonian, as well as all the observables of interest, can be meaningfully expanded in powers of the finite-momentum bosons.
The lowest non-trivial contribution from the finite-momentum modes corresponds to LSW theory, in which the finite-momentum modes are decoupled from the zero-momentum one. Therefore, retaining only this contribution, one obtains a picture of an approximate \emph{separation of variables} between the non-linear quantum-rotor variable (zero-momentum bosons) and the finite-momentum spin waves. 
\end{itemize}
In the following we shall dub our approach the rotor/spin-wave (RSW) theory. We specify our theory to the treatment of XXZ models with power-law decaying interactions. Using quantum Monte Carlo results as benchmark, we show that RSW theory provides a quantitative account of the ground-state physics of the models, in a way similar (and on some accounts superior) to LSW theory with gapped-out zero mode. But our most important result is the description of the \emph{excitation spectrum} and \emph{non-equilibrium dynamics}. Comparing our results with exact diagonalization, we show that RSW is the only spin-wave-based approach that can correctly account for the low-energy excitation spectrum of a \emph{finite-size} system, describing together the Anderson ToS and the spin-wave excitations. The correct description of the low-energy excitation spectrum ensures the ability of the method to describe low-energy \emph{quench dynamics} starting from a fully polarized spin state. Comparing our results with time-dependent variational Monte Carlo based on a pair-product wavefunction, we show that RSW theory is the only spin-wave-based approach that allows for a quantitative description of the dynamics, due to the correct treatment of all the non-linearities of the zero-momentum bosons. A complementary, extensive discussion of the success of the RSW approach in treating quench dynamics is offered by our companion paper, Ref.~\cite{Roscildeetal2023}.  
The quantitative accuracy of our results for the spectral and dynamical properties fundamentally shows that the picture of an approximate separation of variables between a zero-momentum rotor variable and finite-momentum spin-wave ones is a very fruitful playground to understand the behavior of finite-size quantum magnets. 

 Our article is structured as follows. Sec.~\ref{s.spinboson} illustrates the spin-boson mapping and the conventional approach to finite-size spin-wave theory; Sec.~\ref{s.RSW} discusses the approximate rotor/spin-wave separation and RSW theory;  Sec.~\ref{s.equilibrium} compares the predictions of conventional spin-wave theory, RSW theory and quantum Monte Carlo for the ground state of two-dimensional long-range XXZ models;  Sec.~\ref{s.spectrum} discusses the low-energy spectrum; and Sec.~\ref{s.quench} illustrates the problems of conventional spin-wave theory and the success of RSW theory when describing the non-equilibrium dynamics. Conclusions are drawn in Sec.~\ref{s.conclusions}.

\section{Spin-to-boson mapping and spin-wave theory for XXZ models}
\label{s.spinboson}

\subsection{Spin Hamiltonian}
In this work we focus our attention on XXZ models with interactions decaying as a power-law of the distance (hereafter called $\alpha$-XXZ models), 
\begin{equation}
{\cal H}_{\alpha-{\rm XXZ}} = - \sum_{i< j}  J_{ij} \left ( S_i^x S_j^x + S_i^y S_j^y + \Delta S_i^z S_j^z  \right) 
\label{e.alpha-XXZ}
\end{equation}
Here $S_i^\mu$ ($\mu = x,y,z$) are quantum spin operators of arbitrary length, ${\bm S}_i^2 = S(S+1)$; the $i$ and $j$ indices run over the lattice sites (of coordinates ${\bm r}_i$ and ${\bm r}_j$) of a periodic Bravais lattice which is otherwise arbitrary. The hypothesis of a Bravais lattice is not essential and it is only a simplifying one: our treatment can be readily generalized to non-Bravais lattices.
The couplings $J_{ij}$ have a power-law decaying structure with the intersite distance
\begin{equation}
J_{ij} = \frac{J}{|{\bm r}_i - {\bm r}_j|^\alpha}
\end{equation}
with exponent $\alpha\geq 0$; $J>0$ is the ferromagnetic coupling for the $x$ and $y$ spin components; and $\Delta$ is the coupling anisotropy.

In the rest of this work we shall be concerned with systems which, in their ground state, develop long-range ferromagnetic order in the $xy$ plane. This imposes fundamental conditions on the lattice dimensionality as well as on the value of $\Delta$. An easy-plane anisotropy, namely $|\Delta|<1$, guarantees that the $xy$ plane hosts the strongest spin-spin correlations; and the correlation function 
\begin{equation}
C^{(\mu\mu)}_{ij} = \langle S_i^\mu S_j^\mu \rangle  
\end{equation}
is long-ranged in the ground state for $\mu = x$ or $y$ up to a finite critical temperature under the condition that  $\alpha < 2d$ in $d=1,2$, guaranteeing the violation of Mermin-Wagner theorem  \cite{Bruno2001}; and for any $\alpha$ when $d=3$. For any value of $\alpha$ as well, the quantum easy-plane ferromagnet features long-range order in the ground state when $d\geq 2$.  Long-range order for the $x$ and $y$ spin components can also be present for $\Delta < - 1$, namely for a dominant antiferromagnetic interaction of the $z$ spin components, provided that this interaction is sufficiently frustrated by the lattice geometry and/or by the long-range nature of the interactions -- see \emph{e.g.} Ref.~\cite{Frerot2017PRB} for the mean-field phase diagram of the $\alpha$-XXZ model on the square lattice. 

The $\alpha$-XXZ model is not only very relevant for the description of magnetism in the solid state (especially so in the limit of short-range interactions), but it is also implemented (even in its long-range versions) in many platforms of quantum simulation, going from trapped ions \cite{Monroe2021RMP} to Rydberg atoms in optical tweezer arrays \cite{BrowaeysL2020,Chenetal2022}, to ultracold molecules \cite{Christakisetal2023} as well as magnetic atoms \cite{Lepoutreetal2019,Chomazetal2022} in deep optical lattices. 

\subsection{Spin-boson transformation and spin-wave theory}
\label{s.SWtheory}
The crucial step of our approach consists in mapping the spin model onto a bosonic model by using the well-known Holstein-Primakoff (HP) transformation \cite{HP1940}, with quantization axis chosen along the $x$ axis (namely in the $xy$ plane, in which long-range order appears): 
\begin{eqnarray}
S_i^x  & = &  S - n_i \\
S_i^y & = &  \frac{1}{2} \left ( S_i^+ + S_i^- \right) = \frac{1}{2} \left ( \sqrt{2S-n_i} ~b_i + b_i^\dagger\sqrt{2S-n_i} \right ) \nonumber \\
S_i^z & = &  \frac{1}{2i} \left ( S_i^+ - S_i^- \right) = \frac{1}{2i} \left ( \sqrt{2S-n_i} ~b_i  - b_i^\dagger\sqrt{2S-n_i}  \right ) \nonumber~.
\label{e.HP}
\end{eqnarray}
Here the raising and lowering operators, $S_i^+$ and $S_i^-$ respectively are referred to the $x$ axis; and $b_i, b_i^\dagger$ are bosonic operators, with $n_i = b_i^\dagger b_i$.  

If the Hamiltonian of Eq.~\eqref{e.alpha-XXZ} has ferromagnetic long-range order in the $xy$ plane, the mean-field approximation to such a ground state is the coherent spin state (CSS) with all spins aligned along \emph{e.g.} the $x$ axis, $|{\rm CSS}_x\rangle = |\rightarrow_x \rangle^{{\otimes}^N}$, corresponding to the vacuum of the HP bosons.

Under the HP transformation, the $\alpha$-XXZ Hamiltonian takes the form 
\begin{equation}
{\cal H}_{\alpha-{\rm XXZ}} =   - \sum_{i< j}  J_{ij} \left ( {\cal H}_{ij}^{(xx)} + {\cal H}_{ij}^{(++)} + {\cal H}_{ij}^{(+-)} \right) 
\label{e.HPHam}
\end{equation}
where 
\begin{eqnarray}
{\cal H}_{ij}^{(xx)} &= &  (S - n_i)(S-n_j) \nonumber \\ 
 {\cal H}_{ij}^{(++)} &= & \frac{1-\Delta}{4} \left ( \sqrt{2S-n_i} ~b_i \sqrt{2S-n_j} ~b_j + {\rm h.c.}   \right ) \nonumber  \\
  {\cal H}_{ij}^{(+-)} & = & \frac{1+\Delta}{4} \left ( \sqrt{2S-n_i} ~b_i b^\dagger_j \sqrt{2S-n_j}  + {\rm h.c.}   \right )  
\end{eqnarray}

Because of the presence of the square roots in the HP transformation, the above Hamiltonian is highly non-linear. Yet, upon expanding the square roots, it is easy to recognize that it contains only terms which are of even order in the bosonic operators, namely
\begin{equation}
{\cal H}_{\alpha-{\rm XXZ}} =  E_{\rm CSS} + {\cal H}_2 + {\cal H}_4 + ... 
\end{equation}
where 
\begin{equation}
E_{\rm CSS} = -\sum_{i<j} J_{ij} S^2 
\label{e.ECSS}
\end{equation}
is the energy of the CSS (or the mean-field energy), while 
\begin{eqnarray}
&& {\cal H}_2 = -\sum_{i<j} J_{ij} S \Big [ -(n_i + n_j)  \nonumber \\
&& + \frac{1-\Delta}{2} \left ( b_i b_j + b^\dagger_i b^\dagger_j   \right ) 
+ \frac{1+\Delta}{2} \left ( b_i^\dagger b_j + b^\dagger_j b_i   \right ) \Big ] 
\label{e.H2real}
\end{eqnarray}
is the Hamiltonian describing quadratic fluctuations around the mean field, which is at the basis of the spin-wave approximation. 

Introducing the HP bosons in momentum space 
\begin{equation}
b_i = \frac{1}{\sqrt{N}} \sum_{\bm q} e^{i \bm q \cdot \bm r_i} b_{\bm q}~
\end{equation}
where the $\bm q$ wavevectors run over the Brillouin zone of the periodic lattice, one obtains the following form for the quadratic Hamiltonian
\begin{equation}
{\cal H}_2 = \frac{1}{2} \sum_{\bm q} \begin{pmatrix} b_{\bm q}^\dagger \\ b_{-\bm q} \end{pmatrix}^T
\begin{pmatrix} A_{\bm q} & B_{\bm q} \\ B_{\bm q} &  A_{\bm q} \end{pmatrix} 
\begin{pmatrix} b_{\bm q} \\ b^\dagger_{-\bm q} \end{pmatrix} - \frac{1}{2} \sum_{\bm q} A_{\bm q}
\label{e.H2q}
\end{equation}
where 
\begin{eqnarray}
A_{\bm q} & = &  S \left [ J_0 - J_{\bm q}(1+\Delta)/2 \right]  \nonumber \\
B_{\bm q} & =  & - J_{\bm q} S (1-\Delta)/2
\end{eqnarray}
 and where we have introduced the Fourier transform of the spin-spin couplings 
\begin{equation}
J_{\bm q} = \frac{1}{N} \sum_{ij} e^{i{\bm q} \cdot (\bm r_i - \bm r_j)} ~J_{ij} ~.
\end{equation}

The quadratic Hamiltonian can be Bogolyubov-diagonalized by introducing the operators $a_{\bm q}$ and $a^\dagger_{\bm q}$ such that
$b_{\bm q} = u_{\bm q} a_{\bm q} - v_{\bm q} a^\dagger_{-\bm q}$, with 
\begin{equation}
u_{\bm q}  =  \sqrt{\frac{1}{2} \left ( \frac{A_{\bm q}}{\epsilon_{\bm q}} + 1 \right )}~~~
v_{\bm q} = {\rm sign}(B_{\bm q}) \sqrt{\frac{1}{2} \left ( \frac{A_{\bm q}}{\epsilon_{\bm q}} - 1 \right )}
\label{e.Bogo}
\end{equation} 
leading to the form 
\begin{equation}
{\cal H}_2 = \sum_{\bm q} \epsilon_{\bm q} a_{\bm q}^\dagger a_{\bm q} + \frac{1}{2} \sum_{\bm q} (\epsilon_{\bm q} - A_{\bm q})
\end{equation}
where
\begin{equation}
\epsilon_{\bm q} = \sqrt{A_{\bm q}^2 - B_{\bm q}^2} = S \sqrt{(J_0 - J_{\bm q})(J_0 - \Delta J_{\bm q})}~.
\end{equation}
These results form the basis of standard LSW theory. In particular the ground-state energy within this theory is given by 
\begin{equation}
E_0 = E_{\rm CSS} + \frac{1}{2}\sum_{\bm q} (\epsilon_{\bm q} - A_{\bm q})~.
\label{e.E0SW}
\end{equation}

\subsection{Regularization of LSW theory by application of a field (LSW+h approach)}
\label{s.LSWh}

The spin-wave dispersion relation $\epsilon_{\bm q}$ vanishes for ${\bm q} =0$: the existence of this zero mode leads to a singularity in the Bogolyubov transformation of Eq.~\eqref{e.Bogo}, calling for a separate treatment of the $b_0, b_0^\dagger$ operators. 

A possible strategy -- pursued e.g. in Refs.~\cite{Songetal2011,FrerotR2015,Frerot2017PRB} -- to fix the singularity of the Bogolyubov transformation for the zero mode is to gap it out by applying a uniform magnetic field which couples to the order parameter. This implies adding a term ${\cal H}_h  = - h \sum_i S_i^x  = - h SN+h\sum_i n_i$ to the Hamiltonian, which leads to an extra term $- hSN$ in the mean-field energy; and an extra term in the quadratic Hamiltonian, which amounts to a shifted value of the $A_{\bm q}$ coefficient:
\begin{equation}
A_{\bm q} = S \left [ J_0 - J_{\bm q}(1+\Delta)/2 \right] + h~.
\end{equation}
As a consequence a gap appears at ${\bm q} = 0$, $\epsilon_{\bm q=0} = \sqrt{2A_0 h + h^2}$.
The size of the added field is a priori arbitrary: yet a sensible criterion is to choose  $h$ such that the average order parameter is zero in the ground state (or more generally in the equilibrium state) of the system. Denoting with $\langle ... \rangle_h$ the equilibrium averages in the presence of the applied field, one requires that $\langle S_i^x \rangle_h=0$. Given that $\langle  S_i^x \rangle_h = S - N^{-1} \sum_{\bm q} \left [ (u_{\bm K}^2 + v_{\bm K}^2) n_{\bm q} + v^2_{\bm q} \right ]$, where $n_{\bm q} = (e^{\beta \epsilon_{\bm q}}-1)^{-1}$ is the Bose distribution at inverse temperature $\beta = (k_B T)^{-1}$, the condition on $h$ reads 
\begin{equation}
\frac{1}{N} \sum_{\bm q} \frac{ (2n_{\bm q} +1) A_{\bm q}}{2\epsilon_{\bm q}} = S + \frac{1}{2}~.
\end{equation}
This means that the contribution at ${\bm q}=0$ to the sum is not divergent, but leads to a term at most of ${\cal O}(1)$, namely $(2n_0 + 1)A_0/\epsilon_0 \sim {\cal O}(N)$. At low fields such that $\beta \epsilon_0 \ll 1$, one has $n_0 \approx k_B T/ \sqrt{2 A_0 h}$. As a consequence the above condition reads
\begin{equation}
 \frac{\frac{2 k_B T}{\sqrt{2 A_0 h}} + 1}{\sqrt{h}} \sim N~,
\end{equation}
implying that $h\sim {\cal O}(N^{-2})$ at $T=0$, and $h\sim {\cal O}(N^{-1})$ for $T>0$ (and $T > {\cal O}(N^{-1})$). The scaling of the gap at $T=0$ as $\sqrt{h} \sim O(N^{-1})$ interestingly reflects the exact finite-size scaling of the lowest energy excitations in a system which breaks a continuous symmetry in the thermodynamic limit \cite{FrerotR2015}, namely the scaling of the Anderson tower-of-state excitations that we shall discuss below.  
In the following we shall refer to this strategy of regularization of LSW theory as ``LSW+h". 

\section{Zero mode as a quantum rotor, and rotor/spin-wave separation}
\label{s.RSW}

Our strategy to cure the zero-mode problem of LSW theory consists in treating the bosons $b_0, b_0^\dagger$ separately from the finite-momentum ones, in the spirit of Ref.~\cite{ZhongS1993}. As already mentioned, at a technical level this is called for by the singularity of the Bogolyubov transformation, Eq.~\eqref{e.Bogo} for ${\bm q}=0$. Yet the singularity in question is signaling a deep flaw of LSW theory in the presence of gapless modes when applied to finite-size systems. The linearization of the HP transformation, Eq.~\eqref{e.HP}, leading to the quadratic Hamiltonian ${\cal H}_2$, is only valid under the assumption that the gas of HP bosons is dilute, namely $\langle n_i \rangle = N^{-1} \sum_{\bm q} \langle b_{\bm q}^\dagger b_{\bm q} \rangle \ll 2S $ (in a translationally invariant system). This in turn would generally imply $\langle b_{\bm q}^\dagger b_{\bm q} \rangle \ll 2NS$ for all ${\bm q}$'s -- and in fact $\langle b_{\bm q}^\dagger b_{\bm q} \rangle \ll 2S$ for most of the modes. Clearly this assumption cannot hold for ${\bm q}=0$, as $\langle b_0^\dagger b_0 \rangle \sim 1/\epsilon_0 \to \infty$ if the ${\bm q}=0$ mode is gapless. This flaw signals the fundamental fact that the assumption of spontaneous symmetry breaking, and of small quantum fluctuations around a classically ordered state, is untenable on finite-size systems. 

The fact that the population of the ${\bm q=0}$ bosons $\langle b_0^\dagger b_0 \rangle$ cannot be considered as small calls in turn for a treatment of the \emph{non-linear} terms involving the $b_0, b_0^\dagger$ bosons appearing in the bosonic Hamiltonian, Eq.~\eqref{e.HPHam}. Our strategy consists in resumming the non-linear terms including exclusively the zero-momentum bosons \emph{to all orders}, therefore taking their non-linearity fully into account. In so doing, we reconstruct the true nature of the zero-momentum excitations in a finite-size system, which are not linear bosonic quasi-particles, but rather the non-linear excitations of a macroscopic quantum rotor, as we shall illustrate below.

\subsection{Reconstruction of the quantum-rotor variable}
The central insight of our approach consists in the idea that the $b_0, b_0^\dagger$ operators give parametrically larger contributions to the bosonic Hamiltonian than the operators $b_{\bm q \neq 0}, b_{\bm q \neq 0}^\dagger$, as, for all states of interest, $\langle b_0^\dagger b_0 \rangle \gg \langle b_{\bm q \neq 0}^\dagger b_{\bm q \neq 0} \rangle$. This in spirit is similar to the hypothesis of Bose condensation which is the basis of Bogolyubov theory for the diluted Bose gas \cite{PitaevskiiStringari}. Yet, unlike in that theory, we shall not treat the ${\bm q=0}$ bosons via a classical-field approximation, as this would bring us back to the assumption of spontaneous symmetry breaking. On the contrary, we shall fully retain the quantum nature of the ${\bm q}=0$ bosonic mode. 

In view of the presence of a possibly macroscopic number of ${\bm q=0}$ bosons, we shall isolate in the Hamiltonian the part that contains \emph{uniquely} the $b_0, b_0^\dagger$  operators. This amounts to expressing the full bosonic Hamiltonian, Eq.~\eqref{e.HPHam}, in momentum space, and discard all terms containing some $ b_{\bm q \neq 0}, b^\dagger_{\bm q \neq 0}$ operators. Due to the non-linear nature of the Hamiltonian, this may appear as a rather arduous task; yet, to the contrary the task is rather elementary. 

\subsubsection{Zero-momentum/finite-momentum decomposition of operators}

First of all, let us introduce the zero-momentum/finite-momentum decomposition of an operator $O = O(\{ b_i, b_i^\dagger\})$ as
\begin{equation}
O = [O]_{\rm ZM}(b_0,b_0^\dagger) + [O]_{\rm FM}
\end{equation}
where the first (zero-momentum) term contains \emph{uniquely} $b_0,b_0^\dagger$ operators, while the second (finite-momentum, FM) term is a sum of products of bosonic operators containing at least one bosonic operator at finite momentum. For instance, the bosonic operator in real space decomposes as
$b_i = [b_i]_{\rm ZM} + [b_i]_{\rm FM}$ where, quite simply 
\begin{equation}
[b_i]_{\rm ZM} = \frac{b_0}{\sqrt{N}} 
\end{equation}
and
\begin{equation}
[b_i]_{\rm FM} = \frac{1}{\sqrt{N}} \sum_{\bm q \neq 0} e^{i \bm q \cdot \bm r_i} b_{\bm q}~.
\label{e.bFM}
\end{equation}

Let us now move to the spin operators of length $S$, whose $\mu$ $(=x,y,z)$ component is expressed via the bosonic ones as $S_i^\mu = f_{S}^{\mu}(b_i,b_i^\dagger)$, with the functions $f_S^{\mu}$ given by the HP transformation of Eq.~\eqref{e.HP}. 
For those operators, it is immediate to verify the property that 
\begin{equation}
[S_i^{\mu}]_{\rm ZM} =  f_{S}^{\mu}\left( \frac{ b_0}{\sqrt{N}} , \frac{ b_0^\dagger}{\sqrt{N}} \right ) = \frac{f_{NS}^\mu (b_0, b_0^\dagger)}{N} = \frac{K^\mu}{N}
\end{equation}
namely the zero-momentum component of spin-$S$ $S^{\mu}$ operator is equivalent (up to a rescaling factor of $N^{-1}$) to another spin operator $K^\mu$, of macroscopic length $NS$, which is related via the HP transformation  to the $b_0, b_0^\dagger$ operators, namely
\begin{eqnarray}
K^x & = & NS - b_0^\dagger b_0 \nonumber \\
K^y & = &  \frac{1}{2} \left ( \sqrt{2NS - b_0^\dagger b_0} ~b_0 + {\rm h.c.} \right ) \nonumber \\
K^y & = &  \frac{1}{2i} \left ( \sqrt{2NS - b_0^\dagger b_0} ~b_0 - {\rm h.c.} \right )~.
\end{eqnarray}

\subsubsection{Zero-momentum Hamiltonian as a planar-rotor Hamiltonian}
\label{s.ZMHam}

The decomposition of all operators into a zero-momentum part and a finite-momentum one can be readily applied to the $\alpha$-XXZ Hamiltonian, to reconstruct its zero-momentum component
\begin{align}
[{\cal H}_{\alpha-{\rm XXZ}}]_{\rm ZM} =  &
-\frac{1}{2} \sum_{ij} J_{ij}  \Big ( [S_i^x]_{\rm ZM} [S_j^x]_{\rm ZM} + [S_i^y]_{\rm ZM} [S_j^y]_{\rm ZM} \nonumber \\
\phantom{[{\cal H}_{\alpha-{\rm XXZ}}]_{\rm ZM} =  } & ~~~~~~~~~~~~~ +  \Delta [S_i^z]_{\rm ZM} [S_j^z]_{\rm ZM} \Big ) \nonumber \\
= & -\frac{J_{\bm q=0}}{2N}  \left [ (K^x)^2 + (K^y)^2 + \Delta (K^z)^2 \right ]
\label{e.HZM_0}
\end{align}
which, in terms of the macroscopic spin, ${\bm K}  = (K^x, K^y,K^z)$ has the form of a quantum-rotor Hamiltonian. It can be even more explicitly cast in that form 
by using  the fact that $(K^x)^2 + (K^y)^2 = {\bm K}^2 - (K^z)^2$ with  ${\bm K}^2 = NS(NS+1)$, so that the Hamiltonian takes the one-axis-twisting (OAT) \cite{Kitagawa1993PRA} form 
\begin{equation}
[{\cal H}_{\alpha-{\rm XXZ}}]_{\rm ZM}  = {\cal H}_{\rm R} = E_{0,{\rm R}} + \frac{(K^z)^2}{2\tilde{I}}
\label{e.Hrot}
\end{equation}
where
\begin{equation}
E_{0,{\rm R}} = -\frac{S\left (S+\frac{1}{N} \right )}{2}\sum_{ij} J_{ij}
\end{equation}
is the rotor ground-state energy; and the extensive moment of inertia $\tilde{I}$ of the planar rotor is given by 
\begin{equation}
\frac{1}{2\tilde{I}} = \frac{J_{\bm q=0}~(1-\Delta)}{2N}~. 
\label{e.1_2Itilde}
\end{equation}
The moment of inertia becomes negative for $\Delta > 1$. This signals the breakdown of the construction of a bosonic Hamiltonian relying on the HP transformation with quantization axis along $x$, which is obvious when considering that for $\Delta > 1$ the exact ground state of the Hamiltonian is in fact a CSS aligned with the $z$ axis. 

We remark already at this stage that the definition of the rotor Hamiltonian will be further refined (to order ${\cal O}(1/N)$) in Sec.~\ref{s.Dicke}. 
We would also like to stress that the ${\bm K}$ quantum spin is a distinct variable from the collective spin ${\bm J} = \sum_{i=1}^N {\bm S}_i$, whose length ${\bm J}^2$ is not fixed, as the collective spin takes contributions from the finite-momentum bosons as well. Yet their behaviors are strongly related, as we shall see in Sec.~\ref{s.observables}.

\subsection{Finite-momentum bosons: spin-wave Hamiltonian and coupling to the rotor}

As seen in the previous section, the zero-momentum/finite-momentum decomposition of the $\alpha$-XXZ Hamiltonian, 
${\cal H}_{\alpha-{\rm XXZ}} = [{\cal H}_{\alpha-{\rm XXZ}}]_{\rm ZM} + [{\cal H}_{\alpha-{\rm XXZ}}]_{\rm FM}$
leads to the identification of the zero-momentum part as a quantum-rotor Hamiltonian for the macroscopic spin ${\bm K}$. The finite-momentum part, 
$[{\cal H}_{\alpha-{\rm XXZ}}]_{\rm FM}$ is the sum of all terms containing at least one finite-momentum bosonic operator  $b_{\bm q\neq 0}, b^\dagger _{\bm q\neq 0}$, and it describes the energetics of the finite-momentum excitations; as well as their coupling to the zero-momentum ones, namely to the quantum rotor. 
The general structure of $[{\cal H}_{\alpha-{\rm XXZ}}]_{\rm FM}$ reads as
\begin{equation}
[{\cal H}_{\alpha-{\rm XXZ}}]_{\rm FM} =  [{\cal H}_2]_{\rm FM} + [{\cal H}_4]_{\rm FM} + ....
\label{e.HFM}
\end{equation}

The finite-momentum quadratic Hamiltonian corresponds to the spin-wave Hamiltonian \emph{without} the zero mode, namely 
$[{\cal H}_2]_{\rm FM} = {\cal H}_{\rm SW}$ with
\begin{equation}
{\cal H}_{\rm SW} = \frac{1}{2} \sum_{\bm q \neq 0} \begin{pmatrix} b_{\bm q}^\dagger \\ b_{-\bm q} \end{pmatrix}^T
\begin{pmatrix} A_{\bm q} & B_{\bm q} \\ B_{\bm q} &  A_{\bm q} \end{pmatrix} 
\begin{pmatrix} b_{\bm q} \\ b^\dagger_{-\bm q} \end{pmatrix}- \frac{1}{2} \sum_{\bm q \neq 0} A_{\bm q}~,
\label{e.Hsw}
\end{equation}
differing from Eq.~\eqref{e.H2q} by the absence of the ${\bm q=0}$ terms.
Hence this Hamiltonian can be Bogolyubov-diagonalized without any pathology. 

The next-order term, containing the lowest-order non-linearity for the finite-momentum bosons, as well as their coupling to the zero-momentum ones, is represented by the finite-momentum part $[{\cal H}_4]_{\rm FM}$ of the quartic Hamiltonian:
\begin{eqnarray}
&& {\cal H}_4 = -\frac{1}{2} \sum_{ij} J_{ij} \Big [ b_i^\dagger b_j^\dagger b_i b_j \nonumber \\
&& - \frac{1-\Delta}{8} \left ( b_i^\dagger b_i b_i b_j + b_j^\dagger b_j b_j b_i+ {\rm h.c.} \right)   \nonumber \\  
&&- \frac{1+\Delta}{8} \left ( b_i^\dagger b_i b_i b^\dagger_j + b_j^\dagger b^\dagger_j b_j b_i+ {\rm h.c.} \right)  \Big ]~.
\label{e.H4}
\end{eqnarray}
In principle $[{\cal H}_4]_{\rm FM}$ contains terms which are cubic, quadratic, linear and of zero-th order in the zero-momentum bosonic operators $b_0, b_0^\dagger$. Nonetheless, in translationally invariant lattices only momentum-conserving terms are allowed: a term in $[{\cal H}_4]_{\rm FM}$ which is cubic in $b_0, b_0^\dagger$ is clearly not momentum conserving, as the creation/destruction of a boson at finite momentum cannot be momentum-matched by the creation/destruction of zero-momentum ones. Therefore the terms in $[{\cal H}_4]_{\rm FM}$ contaning the highest number of $b_0, b_0^\dagger$ are quadratic in the latter operators. 

\subsection{Approximate rotor/spin-wave  separation} 

In summary, the total Hamiltonian reads 
\begin{equation}
{\cal H}_{\alpha-{\rm XXZ}} = {\cal H}_{\rm R} + {\cal H}_{\rm sw} + [{\cal H}_4]_{\rm FM} + ...
\end{equation}
It is then instructive to compare the order of magnitude of the various terms appearing in the above Hamiltonian. Our treatment rests upon the assumption that, for the low-energy states of interest, or during the non-equilibrium time evolution, finite-momentum bosons form a dilute gas, namely $\langle b_{\bm q\neq 0}^\dagger b_{\bm q \neq 0} \rangle \ll 2S$, such that the operators $b_{\bm q}^{(\dagger)}/\sqrt{2S}$ can be considered as parametrically small. On the other hand, we expect that the population of zero-momentum bosons $\langle b_0^\dagger b_0 \rangle$ can rise up to ${\cal O}(NS)$, when the symmetry of inversion along the quantization ($x$) axis is partially or totally restored. Therefore we have that the operator $b^{(\dagger)}_0/\sqrt{2NS}$ can be considered as larger than $b^{(\dagger)}_{\bm q\neq 0}/\sqrt{2S}$ -- in the sense that, for most of the states of our interest,  $\sqrt{\langle b_{\bm q}^\dagger b_{\bm q} \rangle/(2S)} \lesssim \sqrt{\langle b_0^\dagger b_0 \rangle/(2NS)} \leq 1$. In fact the last inequality is always true, due to constraint on the Hilbert space of the $b_0, b_0^\dagger$  bosons. 

We observe that 
\begin{align}
 &{\cal H}_{\rm R} = (2S)^2N \left [  {\cal O}(1)+ {\cal O}\left ( \frac{b^2_0}{2NS} \right ) +  {\cal O}\left ( \frac{b^4_0}{(2NS)^2} \right )   +... \right ] \nonumber \\
 &{\cal H}_{\rm SW} = (2S)^2N~ {\cal O} \left ( \frac{b_{\bm q\neq 0}^2}{2S} \right )  \nonumber \\
& [{\cal H}_4]_{\rm FM} = (2S)^2N  ~ \Big [ {\cal O} \left ( \frac{b_{\bm q\neq 0}^2}{2S}~ \frac{b^2_0}{2NS} \right ) +   {\cal O} \left ( \frac{b_{\bm q\neq 0}^4}{(2S)^2} \right ) \nonumber \\ 
& ~~~~~~~~~~~~~~~~~~~~~~+ {\cal O} \left ( \frac{b_{\bm q\neq 0}^3}{(2S)^{3/2}}  \frac{b_0}{\sqrt{2NS}} \right ) \Big ]~.
\end{align}

This means that neglecting the coupling terms between the spin waves and rotor, namely neglecting the term $[{\cal H}_4]_{\rm FM}$ and higher-order ones, amounts essentially to neglecting terms of order  ${\cal O} \left ( \frac{b_{\bm q\neq 0}^2}{2S}~ \frac{b^2_0}{2NS} \right )$ (the lowest-order ones in $[{\cal H}_4]_{\rm FM}$ -- see discussion in the previous section) with respect to terms of order ${\cal O} \left ( \frac{b_{\bm q\neq 0}^2}{2S} \right )$ (contained in the spin-wave Hamiltonian). 

This approximation leads to a \emph{separation of variables} between the rotor variable and the finite-momentum linear spin waves. Its justification is physical, descending from the above considerations; as well as technical, as it gives rise to a workable theory, describing harmonic variables coexisting with a quantum-rotor one, all of which are exactly solvable with a moderate computational cost scaling polynomially with $N$. 

\subsection{Improved derivation of zero-momentum/finite-momentum decomposition of operators: projection onto the Dicke-state sector}
\label{s.Dicke}

The above decomposition of the Hamiltonian into zero-momentum and finite momentum components is mathematically very transparent, but it has an immediate drawback: it does \emph{not} reproduce correctly the vacuum expectation value, i.e. the expectation value of the full Hamiltonian on the coherent spin state aligned along $x$,  $\langle {\rm CSS}_x | {\cal H}_{\alpha-{\rm XXZ}} | {\rm CSS}_x\rangle = E_{\rm CSS}$, with $E_{\rm CSS}$  given in Eq.~\eqref{e.ECSS}. 
Indeed, since $|{\rm CSS}_{\rm x}\rangle = |0\rangle$ for the bosons, one immediately finds that, according to the definition of the zero-momentum Hamiltonian in Eq.~\eqref{e.Hrot}:
\begin{equation}
\langle {\rm CSS}_x | [{\cal H}_{\alpha-{\rm XXZ}} ]_{\rm ZM} |{\rm CSS}_x\rangle  = E_{\rm CSS} - \frac{S}{4N} \sum_{ij} J_{ij} (1+2\Delta)
\end{equation}
while $\langle {\rm CSS}_x | {\cal H}_2 |{\rm CSS}_x\rangle=0$, because the quadratic Hamiltonian, Eq.~\eqref{e.H2real}, is normal-ordered. Hence the correct expectation value is missed by an error of order ${\cal O}(1/N)$ compared to the correct term. 

 The origin of this problem is rather clear: if one were to normal-order the whole finite-momentum component of the Hamiltonian, Eq.~\eqref{e.HFM}, this would produce terms (of relative order 
${\cal O}(1/N)$) that would add to the zero-momentum Hamiltonian. This is already apparent in the quartic Hamiltonian given by Eq.~\eqref{e.H4}, which is clearly not normal-ordered, and which, under normal ordering and Fourier transformation, would contain terms involving exclusively the $b_0$, $b_0^\dagger$ operators. Therefore the correct prescription for the zero-momentum/finite-momentum decomposition of the Hamiltonian is 
\begin{equation}
[{\cal H}_{\alpha-{\rm XXZ}}]_{\rm ZM} = {\cal H}_{\alpha-{\rm XXZ}} - :[{\cal H}_{\alpha-{\rm XXZ}}]_{\rm FM}:
\end{equation}
where $:(...):$ indicates normal ordering for the bosonic operators. This may suggest that, in order to reconstruct correctly the zero-momentum Hamiltonian, one should examine (and normal-order) the whole series of terms which have been neglected in the definition of the finite-momentum Hamiltonian, Eq.~\eqref{e.HFM} -- a rather arduous task. 

Yet, luckily, one can get around this difficulty, and fully eliminate all the the problems at order ${\cal O}(1/N)$ by realizing that a fundamental property of the zero-momentum Hamiltonian is that it is fully \emph{symmetric} under permutation of sites, because so are by construction the $b_0, b_0^\dagger$ operators. Moreover the $b_0, b_0^\dagger$ operators live on a Hilbert space of dimensions $2NS+1$ (because of the constraint of the bosonic occupation coming from the physics of the spins). Therefore the correct identification of the zero-momentum Hamiltonian is that of the projection of ${\cal H}_{\alpha-{\rm XXZ}}$ on the $(2NS+1)$-dimensional sector of Hilbert space spanned by the symmetric Dicke states $|J_{\rm tot}=NS,M\rangle$. Introducing the total spin 
\begin{equation}
{\bm J} = \sum_{i=1}^N {\bm S}_i~,
\end{equation}
the Dicke states are eigenstates of ${\bm J}^2$ and $J^z$ with eigenvalues $J_{\rm tot}(J_{\rm tot}+1)$ and $M$ respectively, where $J_{\rm tot}$ takes its maximum value $J_{\rm tot}=NS$; and $M = -J_{\rm tot}, ..., J_{\rm tot}$. 

Therefore we redefine the zero-momentum component of a generic operator $O$ as 
\begin{equation}
[O]_{\rm ZM} = \sum_{M,M'} \langle J_{\rm tot},M|O |J_{\rm tot},M' \rangle ~ |J_{\rm tot},M \rangle \langle J_{\rm tot},M'|~.
\label{e.OZM}
\end{equation}
This definition, along with the definition of the finite-momentum component as being normal-ordered, leads to the general decomposition
\begin{equation}
O = [O]_{\rm ZM} + :[O]_{\rm FM}: 
\label{e.decomposition}
\end{equation}
which guarantees that the expectation value on the CSS are correct, since: 1) the CSS lives in the Dicke subspace; and 2) the normal ordering of the FM component ensures the vanishing of the FM contribution to the expectation value on the CSS. 

The zero-momentum operator of  Eq.~\eqref{e.OZM} can then be expressed as a function of the components $K^\mu$ for a spin of length $NS$, which can be simply viewed as tools to express all operators acting on the Dicke subspace, namely
\begin{align}
K^\mu &  = [J^\mu]_{\rm ZM}  \\ 
& = \sum_{M,M'} \langle J_{\rm tot},M| \sum_i S_i^\mu |J_{\rm tot},M' \rangle ~ |J_{\rm tot},M \rangle \langle J_{\rm tot},M'|~. \nonumber
\end{align}
Therefore we can write 
\begin{equation}
 [S_i^\mu]_{\rm ZM} = \frac{K^\mu}{N} 
 \end{equation}
 and, for $i \neq j$ 
 \begin{equation}
 [S_i^\mu S_j^\nu]_{\rm ZM} = \frac{1}{N(N-1)} \left ( K^\mu K^\nu - \sum_l [S_l^\mu S_l^\nu]_{\rm ZM} \right )~
 \label{e.SiSjZM}
 \end{equation}
 where
 $[S_l^\mu S_l^\nu]_{\rm ZM}$ is again defined as in Eq.~\eqref{e.OZM}. 

The U(1) symmetry of the Hamiltonian makes it diagonal on the Dicke subspace, so that
\begin{align}
& [{\cal H}_{\alpha-{\rm XXZ}}]_{\rm ZM}  = \sum_{M} \langle J,M|{\cal H}_{\alpha-{\rm XXZ}}  |J,M \rangle ~ |J,M \rangle \langle J,M| \nonumber \\
& = \frac{J_0}{2(N-1)}  \big [ - N(N-1)S^2 + (1-\Delta) (K^z)^2  \nonumber \\
&~~~~~~~~~~~~~~~~~~  - (1-\Delta) [(S_i^z)^2]_{\rm ZM} \big ]  \nonumber \\
& = E_{\rm CSS} - \frac{(1-\Delta)}{N-1}~ [(S_i^z)^2]_{\rm ZM} + \frac{(K^z)^2}{2I} = {\cal H}_{\rm R}
\label{e.HZM}
\end{align}
where we have introduced the refined definition of the moment of inertia
\begin{equation}
\frac{1}{2I} = \frac{J_0(1-\Delta)}{2(N-1)}
\end{equation}
differing from the previous definition, Eq.~\eqref{e.1_2Itilde}, by corrections of order ${\cal O}(1/N)$. 
The last line of Eq.~\eqref{e.HZM} defines the rotor Hamiltonian ${\cal H}_{\rm R}$, and it replaces the previous definition of Eq.~\eqref{e.HZM_0}. 
In particular, since $\langle {\rm CSS}_x |  [(S_i^z)^2]_{\rm ZM} | {\rm CSS}_x \rangle = \langle {\rm CSS}_x |  (K^z)^2 | {\rm CSS}_x \rangle/N = S/2$, we obtain that 
$\langle {\rm CSS}_x | [{\cal H}_{\alpha-{\rm XXZ}}]_{\rm ZM} | {\rm CSS}_x \rangle = E_{\rm CSS}$, in agreement with what mentioned above.    

Calculating $[(S_i^z)^2]_{\rm ZM}$ is trivial for $S=1/2$, since in that case one has simply that  $[(S_i^z)^2]_{\rm ZM}=1/4$. On the other hand, it is not trivial at all for $S>1/2$ spins. Since the results of this work (presented in Secs.~\ref{s.equilibrium} and \ref{s.quench}) focus on $S=1/2$ spins, we shall postpone this problem to future work.  

\subsection{RSW separation and Hilbert-space extension}
\label{s.Hilbert}

In the previous subsection we described how one can identify the zero-momentum degrees of freedom as resulting from a projection of the Hamiltonian, as well as of any other operator, onto the Dicke subspace of states with maximum collective spin length. 
The rotor/spin-wave separation at the heart of RSW theory involves then the approximate separation between a rotor variable which lives in the Dicke subspace, and spin-wave degrees of freedom, which describe in turn the projection of the state of the system onto the sectors which are orthogonal to the Dicke subspace. This construction clearly introduces an extension to the mathematical structure of the Hilbert space $H$  of the many-body spin system. Indeed, $H$ is the \emph{direct sum} of orthogonal subspaces with different collective-spin length $J_{\rm tot}$, $H_{J_{\rm tot}}$
\begin{equation}
H = \oplus_{_{J_{\rm tot}}} H_{J_{\rm tot}}~.
\end{equation}
On the other hand, by assuming a separation of variables between variables living in $H_{NS}$ and variables living in the orthogonal subspaces, we are tacitly assuming that the Hilbert space is extended to a \emph{tensor-product} structure:
\begin{equation}
H  \rightarrow H_{NS} \otimes \left ( \oplus_{_{J_{\rm tot}<NS}} H_{J_{\rm tot}}~ \right)~.
\end{equation}
Moreover, by linearizing the Hamiltonian for finite-momentum bosons, the orthogonal subspaces are approximated as an (infinite dimensional) bosonic subspace, leading to further extension -- which is anyway at the heart of most bosonization approaches.   
Embedding a specific problem into a larger space is a rather typical step in order to introduce approximations -- see for instance conventional spin-wave theory, slave-boson approaches \cite{Fresard2015}, Schwinger-boson approaches \cite{Auerbachbook}, etc. In the case of RSW theory, this has minor consequences when evaluating standard observables, because, as we will see, such observables do admit an additive structure in terms of rotor/spin-wave variables when neglecting terms of the same kind as those neglected in the Hamiltonian. The consequences are slightly more serious when evaluating instead entanglement entropies, which, as we will see in Sec.~\ref{s.entropy}, are going to be overestimated in some cases, as an immediate consequence of the enlargement of Hilbert-space dimensions. We shall comment on this aspect further.



\subsection{Observables under rotor/spin-wave separation} 
\label{s.observables}

The zero-momentum/finite-momentum decomposition defined in Sec.~\ref{s.Dicke}, Eq.~\eqref{e.decomposition} can be applied systematically to all observables of interest. 
Within the framework of RSW theory, the FM part is treated in an approximate manner, by retaining only quadratic terms in the finite-momentum bosons and neglecting all couplings to the zero-momentum ones.

\subsubsection{Hamiltonian and its spectrum}

The approximate separation of variables between zero-momentum ones (rotor) and finite-momentum ones (spin waves) defines the additive structure for the Hamiltonian
\begin{equation}
{\cal H}_{\alpha-{\rm XXZ}} \approx {\cal H}_{\rm R} + {\cal H}_{\rm SW} ~,
\end{equation}
where the rotor Hamiltonian is defined as in Eq.~\eqref{e.HZM}. 
The spin-wave Hamiltonian for finite-momentum bosons takes the following  form  after Bogolyubov diagonalization 
 \begin{equation}
{\cal H}_{\rm SW} = \sum_{\bm q\neq 0} \epsilon_{\bm q} a_{\bm q}^\dagger a_{\bm q} + \frac{1}{2} \sum_{\bm q \neq 0} (\epsilon_{\bm q} - A_{\bm q})
\end{equation}
with eigenvectors $|\{n^{(a)}_{\bm q} \}\rangle$ corresponding to Fock states for the Bogolyubov ($a$) quasiparticles. 

The spectrum of the Hamiltonian under RSW theory takes therefore the form 
\begin{equation}
E_{\rm RSW}(M_K,\{n^{(a)}_{\bm q} \}) = E_{\rm gs} + \frac{M_K^2}{2I} + \sum_{\bm q\neq 0} \epsilon_{\bm q} n^{(a)}_{\bm q}  
\label{e.RSWspectrum}
\end{equation}
with $M_K = -NS,..., NS$. The associated approximate eigenstates possess the factorized rotor/spin-wave form $|K=NS,M_K\rangle \otimes |\{n^{(a)}_{\bm q} \}\rangle$ -- see Sec.~\ref{s.Hilbert}.
$E_{\rm gs}$ is the  RSW ground-state energy, whose expression reads
\begin{equation}
 E_{\rm gs} =  E_{\rm CSS} - \frac{J_0(1-\Delta)}{2(N-1)}~ [(S_i^z)^2]_{\rm ZM} + \frac{1}{2} \sum_{\bm q \neq 0} (\epsilon_{\bm q} - A_{\bm q})~.
\end{equation}
This contains the zero-point energy of the rotor (corresponding to the mean-field energy for large $N$), as well as the zero-point energy of the finite-momentum modes. 
This expression nearly coincides with that of LSW+h,  Eq.~\eqref{e.E0SW}, with the second term on the right-hand side playing the role of the $(\epsilon_0 - A_0)/2$ term, removed from the last sum within RSW. Please notice that the actual spectrum calculated in Sec.~\ref{s.spectrum} will differ slightly with respect to the one described here because of a deformation of the rotor Hamiltonian aimed at obtaining a vanishing average magnetization in the ground state, as detailed in Sec.~\ref{s.modification}.   

Here and in the following we denote with $\langle ... \rangle_{\rm R}$  the average over the (equilibrium or non-equilibrium) state of the rotor, and with $\langle ... \rangle_{\rm SW}$ the average over the state of the spin waves. Hence the corresponding average energy takes the form $\langle \cal H \rangle = \langle {\cal H}_{\rm R} \rangle_{\rm R} + \langle {\cal H}_{\rm SW} \rangle_{\rm SW}$. 

\subsubsection{Average total spin}

The average $x$-component of the total spin ${\bm J}$ is given by 
\begin{equation}
\langle J^x \rangle = NS - \sum_{\bm q} \langle b_{\bm q}^\dagger b_{\bm q} \rangle
\end{equation}
while the other two components systematically vanish in all the cases that we shall consider below. 
Within RSW theory, the above average acquires the additive form 
\begin{equation}
\langle J^x \rangle = \langle K^x \rangle_{\rm R} -  \sum_{\bm q\neq 0} \langle b_{\bm q}^\dagger b_{\bm q} \rangle_{\rm SW}~. 
\label{e.Jx}
\end{equation}

\subsubsection{Correlation functions and total-spin covariance matrix}


We consider generic spin-spin correlation functions
 \begin{align}
C^{\mu\nu}_{ij} & = \frac{1}{2} \langle \{ S^{\mu}_i  , S^{\nu}_j \} \rangle  \nonumber \\
& = \frac{1}{2N} \sum_{\bm q} e^{i\bm q \cdot (\bm r_i - \bm r_j)} \langle \{ S_{\bm q}^\mu, S_{-\bm q}^\nu \} \rangle
\end{align}
where we have introduced the  Fourier decomposition of spin operators
\begin{equation}
S^\mu_i = N^{-1/2} \sum_{\bm q} e^{i\bm q\cdot \bm r_i} S^\mu_{\bm q}~.
\end{equation}

Using the prescription of Eqs.~\eqref{e.decomposition} and \eqref{e.SiSjZM}, the correlation function is decomposed as
\begin{align}
C^{\mu\nu}_{ij} & \approx  \frac{1}{2} \langle [\{S_i^\mu ,S_i^\nu \}]_{\rm ZM}\rangle_{\rm R} ~\delta_{ij} \nonumber \\
& +  \frac{\langle \{ K^\mu, K^\nu\}\rangle_{\rm R} - \sum_l \langle [\{S_l^\mu ,S_l^\nu \}]_{\rm ZM}\rangle_{\rm R}}{2N(N-1)} (1-\delta_{ij}) 
  \nonumber \\
& + \frac{1}{2N} \sum_{\bm q \neq 0} e^{i\bm q \cdot (\bm r_i - \bm r_j)} \langle :\{ S_{\bm q}^\mu, S_{-\bm q}^\nu \}: \rangle_{\rm SW}
\label{e.correlations}
\end{align}
where the term $\langle :\{ S_{\bm q}^\mu, S_{-\bm q}^\nu \}: \rangle_{\rm SW}$ is to be understood as the expectation value on the SW state of the (normal-ordered) harmonic approximation to the $\{ S_{\bm q}^\mu, S_{-\bm q}^\nu \}$ operator when expressed in terms of HP bosons. 
The detailed expression of the correlation functions for $\mu,\nu= x,y,z$ is provided in App.~\ref{a.correlations}.


The covariance matrix elements for the collective spin, obtained by integrating the expressions in App.~\ref{a.correlations} over space, take then the form 
\begin{align}
{\rm Var}(J^x)  & \approx {\rm Var}(K^x)_{\rm R} - 2(NS-\langle K^x \rangle_{\rm R})N_{\rm FM} - N^2_{\rm FM} \nonumber \\
{\rm Var}(J^y)  & \approx  {\rm Var}(K^y)_{\rm R} \nonumber  \\
{\rm Var}(J^z)  & \approx  {\rm Var}(K^z)_{\rm R} \nonumber \\
{\rm Cov} (J^x, J^y) & \approx \frac{1}{2} \langle \{ K^x, K^y \} \rangle_{\rm R} \nonumber \\
{\rm Cov} (J^x, J^z)   & \approx \frac{1}{2}\langle\{ K^x, K^z \} \rangle_{\rm R} \nonumber \\
{\rm Cov} (J^y, J^z)   & \approx \frac{1}{2}\langle\{ K^y, K^z \} \rangle_{\rm R} 
\end{align}
where ${\rm Var}(J^\alpha) = \langle (J^\alpha)^2 \rangle - \langle J^\alpha \rangle^2$ is the variance, ${\rm Cov}(J^\alpha, J^\beta) = \frac{1}{2} \langle \{ J^\alpha,J^\beta \} \rangle - \langle J^\alpha \rangle \langle J^\beta \rangle $ is the covariance, and   
$N_{\rm FM} = \sum_{\bm q\neq 0} \langle b^\dagger_{\bm q} b_{\bm q} \rangle_{\rm SW}$ is the population of finite-momentum bosons. 

From the above expression we observe that, within RSW theory for a translationally invariant system with average spin $\langle \bm J \rangle = (\langle J^x \rangle, 0, 0)$, the covariance matrix of the collective spin coincides with that of the rotor -- apart from the term ${\rm Var}(J^x)$ which also receives a contribution from the spin waves.   

\subsubsection{Entanglement entropy}

As already mentioned above, physical states within RSW theory factorize into rotor and spin-wave part:  $|\psi\rangle = |\psi_{\rm R}\rangle \otimes |\psi_{\rm SW}\rangle$. As a consequence, the entanglement entropy of a subsystem has an additive structure, with a contribution from the rotor variable and one from the finite-momentum spin waves. 

The contribution from the spin waves can be calculated from the knowledge of the covariance matrix of the finite-momentum bosons, namely from the matrices $G_{ij} = \langle \tilde{b}^\dagger_i \tilde{b}_j \rangle$ and $F_{ij} = \langle \tilde{b}_i \tilde{b}_j \rangle$, where
the $\tilde{b}_i, \tilde{b}^\dagger_i$ operators only contain finite-momentum components $\tilde{b}_i = [b_i]_{\rm FM}$ as defined in Eq.~\eqref{e.bFM}.  
As a consequence  
\begin{align}
G_{ij} = &\frac{1}{N} \sum_{\bm q \neq 0}  e^{-i \bm q \cdot (\bm r_i-\bm r_j)} \langle  b^{\dagger}_{\bm q} b_{\bm q} \rangle \nonumber \\
F_{ij} = &\frac{1}{N} \sum_{\bm q \neq 0}  e^{i \bm q \cdot (\bm r_i-\bm r_j)} \langle  b_{\bm q} b_{-\bm q} \rangle \nonumber~.
\end{align}
Strictly speaking,  the $\tilde b_i, \tilde b_i^\dagger$ operators satisfy a slightly modified bosonic commutation relation 
\begin{equation}
[\tilde b_i ,\tilde b_j^{\dagger}] = \delta_{ij} \left ( 1 - \frac{1}{N} \right)
\end{equation}
where the $-1/N$ correction comes from the absence of the ${\bm q}=0$ component. Yet for $N \gg 1$ we can safely ignore this aspect and treat the $\tilde b, \tilde b^\dagger$ operators as regular bosonic ones. 

Since the physical states of the finite-momentum bosons are Gaussian states, the reduced density matrix of any subsystem $A$  (comprising $N_A$ sites) is also a Gaussian state, namely the exponential of a quadratic form of the  $\tilde{b}_i, \tilde{b}^\dagger_i$ operators, fully reconstructed from the knowledge of the $G_A$ and $F_A$ matrices, which are the $G$ and $F$ matrices with indices restricted to $i, j \in A$. The one-body eigenfrequencies of the quadratic form defining the reduced state of the subsystem are obtained from the diagonalization of the $2N_A \times 2N_A$ matrix  \cite{FrerotR2015}
\begin{equation}
\begin{pmatrix}  -\mathbb{1}_A- G^*_A   & F_{A}  \\ F^*_{A} & G_A \end{pmatrix}
\end{equation}
whose diagonal form reads ${\rm diag}(\{-1-n_\alpha \},\{ n_{\alpha} \})$ where $\alpha = 1,..., N_A$ and $n_{\alpha} \geq 0$. The entanglement entropy (von Neumann and second R\'enyi, respectively) is then obtained as
\begin{align}
S^{\rm (vN)}_{A,{\rm SW}} & = -\sum_\alpha \left [ n_\alpha \log (n_\alpha) - (1+n_{\alpha}) \log(1+n_{\alpha}) \right ]  \nonumber \\
S^{(2)}_{A,{\rm SW}}  & = \sum_\alpha \log(1+2 n_{\alpha}) ~.
\end{align}

According to the construction of Sec.~\ref{s.Dicke}, the entanglement contribution from the rotor degree of freedom is the entropy of a subsystem $A$ of $N_A$ spins of length $S$ which are involved in a $N$-spin collective symmetric spin state of maximum total spin length $J_{\rm tot}=NS$; or, equivalently, of a subsystem of $2SN_A$ spins of length $S=1/2$ within a system of $2NS$ spins in a symmetric spin state with the same total spin length. The generic state of such a system is a superposition of Dicke states $|J,M\rangle$ 
\begin{equation}
|\psi_{\rm R}\rangle = \sum_{M=-J_{\rm tot}}^{J_{\rm tot}} c_{J_{\rm tot},M} |J,M\rangle
\label{e.Dicke}
\end{equation}
which in turn admit a Schmidt decomposition into sub-system Dicke states \cite{Latorreetal2005}
\begin{equation}
|J_{\rm tot},M\rangle = \sum_{n_A=0}^{2J_A} \sqrt{p_{n_A}}~ |J_A, n_A-J_A\rangle |J_B, M-n_A-J_A\rangle 
\label{e.Schmidt}
\end{equation}
where $J_A = N_A S$, $J_B = (N-N_A)S$,  
\begin{equation}
p_{n_A} = \begin{pmatrix} 2J_A  \\  n_A \end{pmatrix} \begin{pmatrix}  2J_B \\ n - n_A \end{pmatrix} / \begin{pmatrix}  2J_{\rm tot} \\ n\end{pmatrix} 
\end{equation}
and $n = M + J_{\rm tot}$.

The reduced density matrix of subsystem $A$ is readily built as a $(2 J_A +1) \times (2 J_A +1)$ matrix from the Schmidt decomposition of the state 
$|\psi_{\rm R}\rangle$ which results from the combination of Eqs.~\eqref{e.Dicke} and \eqref{e.Schmidt}. Its diagonalization leads then to the entanglement entropy.

The resulting entanglement entropy for the $A$ subsystem is then estimated within RSW theory as
\begin{equation}
S_A \approx S_{A,\rm R} + S_{A,\rm SW}~.
\label{e.entRSW}
\end{equation} 
Because of the extension of the Hilbert space implicit in the RSW theory (see Sec.~\ref{s.Hilbert}), we generally expect this entropy to overestimate the actual entanglement entropy of the state of interest. 
 
\section{Ground-state properties}
\label{s.equilibrium}

In this section we discuss the predictions of RSW theory for the ground-state properties of U(1) symmetric systems. Throughout the rest of this work we shall specialize our attention to the long-range XX model (hereafter referred to as the $\alpha$-XX model), namely Eq.~\eqref{e.alpha-XXZ} with $\Delta=0$ and a variable $\alpha$ exponent. In particular we will concentrate on the case of a square-lattice geometry with $N=L\times L$ spins, guaranteeing that the ground state of the system exhibits long-range order for all values of $\alpha$. 

Conventional LSW theory can be applied as well to the equilibrium physics of this model in the thermodynamic limit; and its regularized version (LSW+h) allows for the treatment of finite-size effects. In the following we shall conduct a systematic comparison of RSW theory with LSW+h one, as well as with quantum Monte Carlo (QMC) results obtained via the Stochastic Series Expansion approach \cite{Syljuasen2002PRE}. 

\subsection{Hamiltonian modification to set the order parameter to zero}
\label{s.modification}
The exact equilibrium state of a finite-size system does not exhibit spontaneous symmetry breaking, namely the order parameter $m = \langle J^x \rangle/N$ is strictly zero at any temperature. In the RSW expression for the order parameter, Eq.~\eqref{e.Jx}, one has that $m = m_{\rm R} + \delta m_{\rm SW}$  with $m_{\rm R} = \langle K^x \rangle_{\rm R}=0$ in thermal equilibrium for the rotor Hamiltonian Eq.~\eqref{e.Hrot}, given its U(1) symmetry.  On the other hand one has in general that $\delta m_{\rm SW} = - N^{-1} \sum_{\bm q\neq 0} \langle b_{\bm q}^\dagger b_{\bm q} \rangle_{\rm SW} < 0$, because of the finite population of FM bosons in any equilibrium state of the system, including the ground state. This means that RSW theory would naively predict $m < 0$, which is unphysical. This results reveals the fact that it is impossible to build a quadratic theory of elementary excitations around a ground state which is fully symmetric under U(1) rotations. 

This apparent problem with RSW theory can be easily corrected for by slightly modifying the rotor Hamiltonian
\begin{equation}
{\cal H}_{\rm R}  \to {\cal H}_{\rm R} - h K^x
\label{e.Hrotfield}
\end{equation}
with the addition of a finite field $h$, such that the rotor contribution to the order parameter  $m_{\rm R}$ becomes finite, and it compensates the SW one, resulting in $m=0$. Given that the rotor spectrum is made of a tower of state separated by energies of order ${\cal O}(1/N)$, a field scaling as $h \approx h_0/N$ is \emph{a priori} sufficient to induce a macroscopic magnetization in it by admixing low-lying Hamiltonian eigenstates. In practice we observe that $h_0 \lesssim 10^{-3} J$ for the models of interest to this work.

We shall make use of this slight modification of the rotor Hamiltonian only in the equilibrium calculations. In the non-equilibrium ones the original U(1) symmetric Hamiltonian shall be used instead.  


\begin{figure}[ht!]
\begin{center}
\includegraphics[width=0.95\columnwidth]{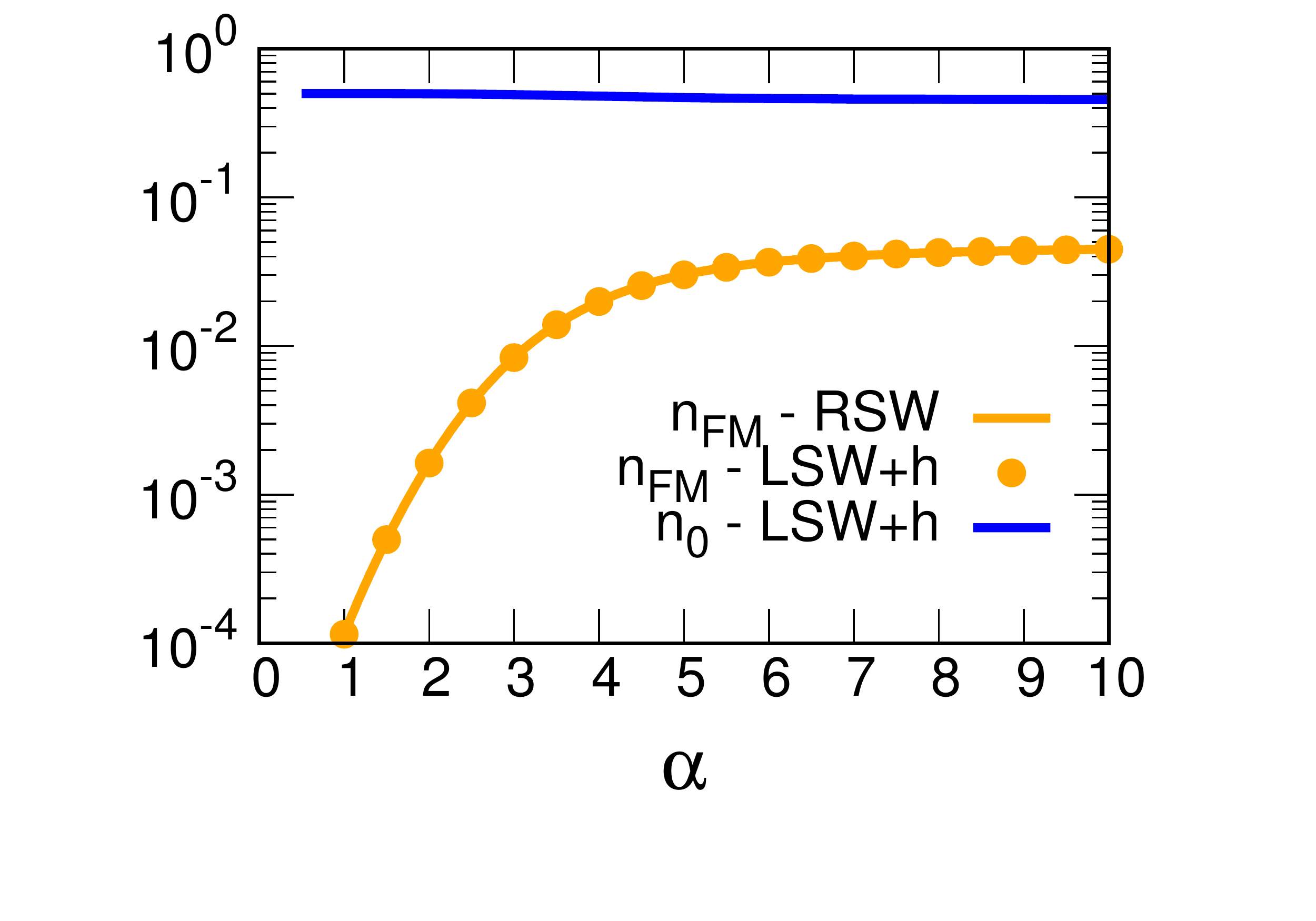}
\caption{Density of finite momentum bosons ($n_{\rm FM}$) and zero-momentum ones ($n_0$) in the ground state of the $S=1/2$ 2$d$ $\alpha-XX$ model with variable $\alpha$. Here we show the predictions of both RSW and LSW+h theory. The calculation has been performed for a $L\times L$ lattice with $L=20$. }
\label{f.nboseGS}
\end{center}
\end{figure}

\subsection{Exactness of RSW theory for $\alpha \to 0$}

A fundamental remark concerns the small-$\alpha$ limit. In this limit, the SW contribution to all quantities vanishes within RSW, as it can be see in Fig.~\ref{f.nboseGS}. There we plot the density of finite-momentum bosons 
\begin{equation}
n_{\rm FM} = \frac{1}{N} \sum_{{\bm q} \neq 0} \langle b^\dagger_{\bm q} b_{\bm q} \rangle 
\end{equation}
which is seen to vanish as $\alpha\to 0$. This implies that in this limit RSW theory recovers the physics of a planar rotor for all system sizes, namely the exact description of the $\alpha=0$ limit. This property is not shared with spin-wave theory in its finite-size formulation, namely LSW+h. While the finite-momentum bosons disappear upon decreasing $\alpha$ (their value is nearly identical to that of RSW theory, as shown in Fig.~\ref{f.nboseGS}), the only degree of freedom that remains active is the $\bm q=0$ mode, whose population alone must satisfy the condition of a vanishing order parameter, $\langle b_0^\dagger b_0 \rangle \to  NS$. This implies that, in the limit $\alpha\to0$, the ground state of the rotor Hamiltonian ($\alpha=0$), which is exactly a Dicke state $|J,M\rangle = |NS,0\rangle$, is approximated within LSW+h theory by a state of a single bosonic mode with a macroscopic population, corresponding to a density $n_0 = \langle b_0^\dagger b_0 \rangle/N \to S$ -- see Fig.~\ref{f.nboseGS}. It remains rather surprising that the linearization of such a mode, at the basis of LSW+h theory, can lead at all to quantitative predictions, for any value of $\alpha$.  Moreover the LSW+h ground state state breaks explicitly the U(1) symmetry of rotation around the $z$ axis, while this symmetry is instead recovered exactly by construction within RSW theory when $\alpha \to 0$; and it is nearly respected for small $\alpha$ (modulo the small symmetry-breaking field $h$ discussed in the previous section).

\begin{figure}[ht!]
\begin{center}
\includegraphics[width=\columnwidth]{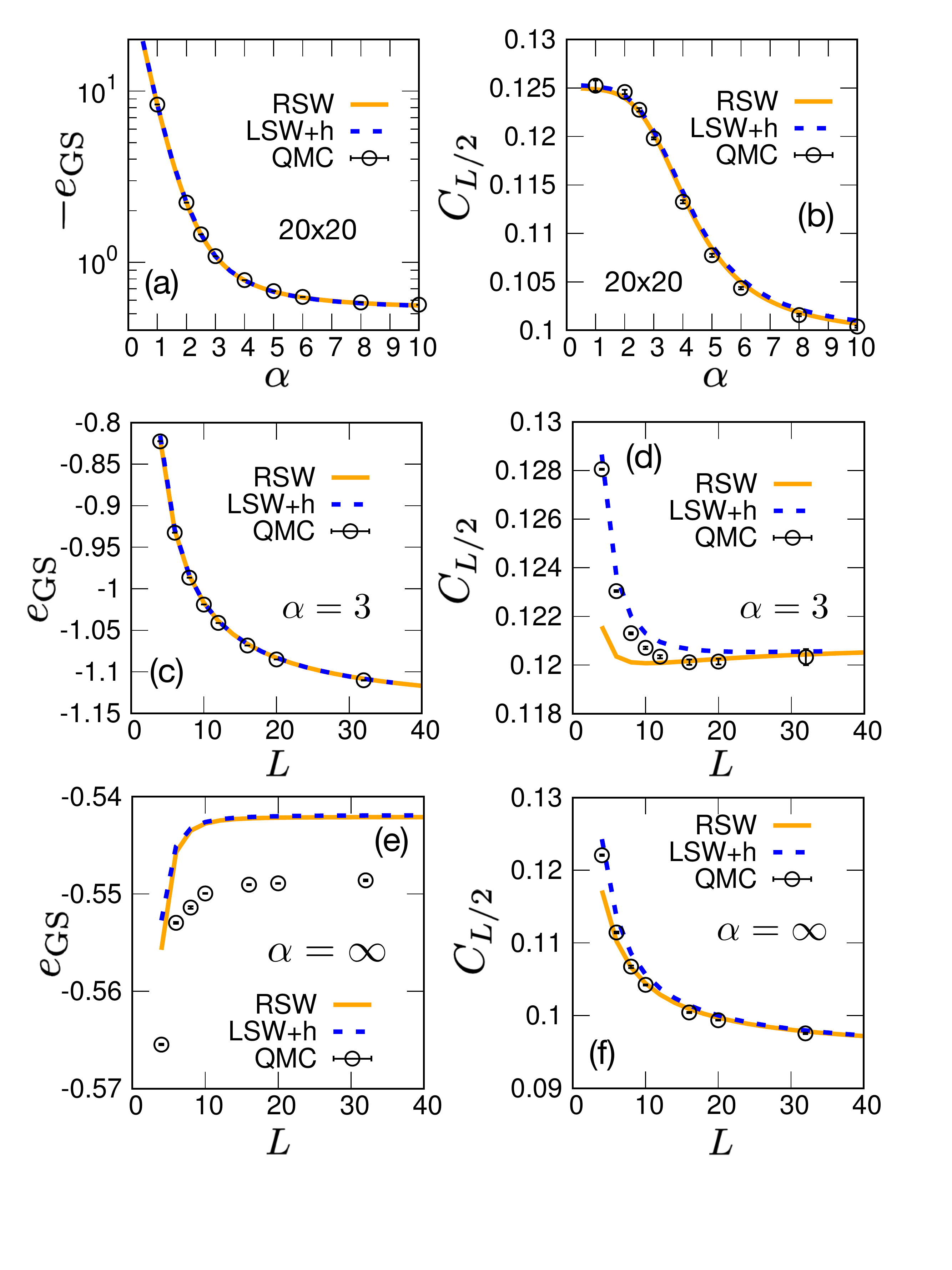}
\caption{\emph{Ground-state properties of the 2$d$ $\alpha-XX$ model.} (a-b) Ground-state energy per spin $e_{\rm gs}$ (a) and correlations $C_{L/2}$ (b) for a $20\times 20$ lattice as a function of $\alpha$. Here and all the other panels we compare the predictions of RSW and LSW+h theory with the numerically exact results from QMC -- in particular, for LSW+h theory $e_{\rm gs} = E_0/N$; (c-d) Size dependence of $e_{\rm gs}$ and $C_{L/2}$ for the dipolar $\alpha-XX$ model ($\alpha=3$); (e-f) same as in (c-d) but for the nearest-neighbor interacting model ($\alpha = \infty$).}
\label{f.equilibrium}
\end{center}
\end{figure}

\begin{figure}[ht!]
\begin{center}
\includegraphics[width=0.95 \columnwidth]{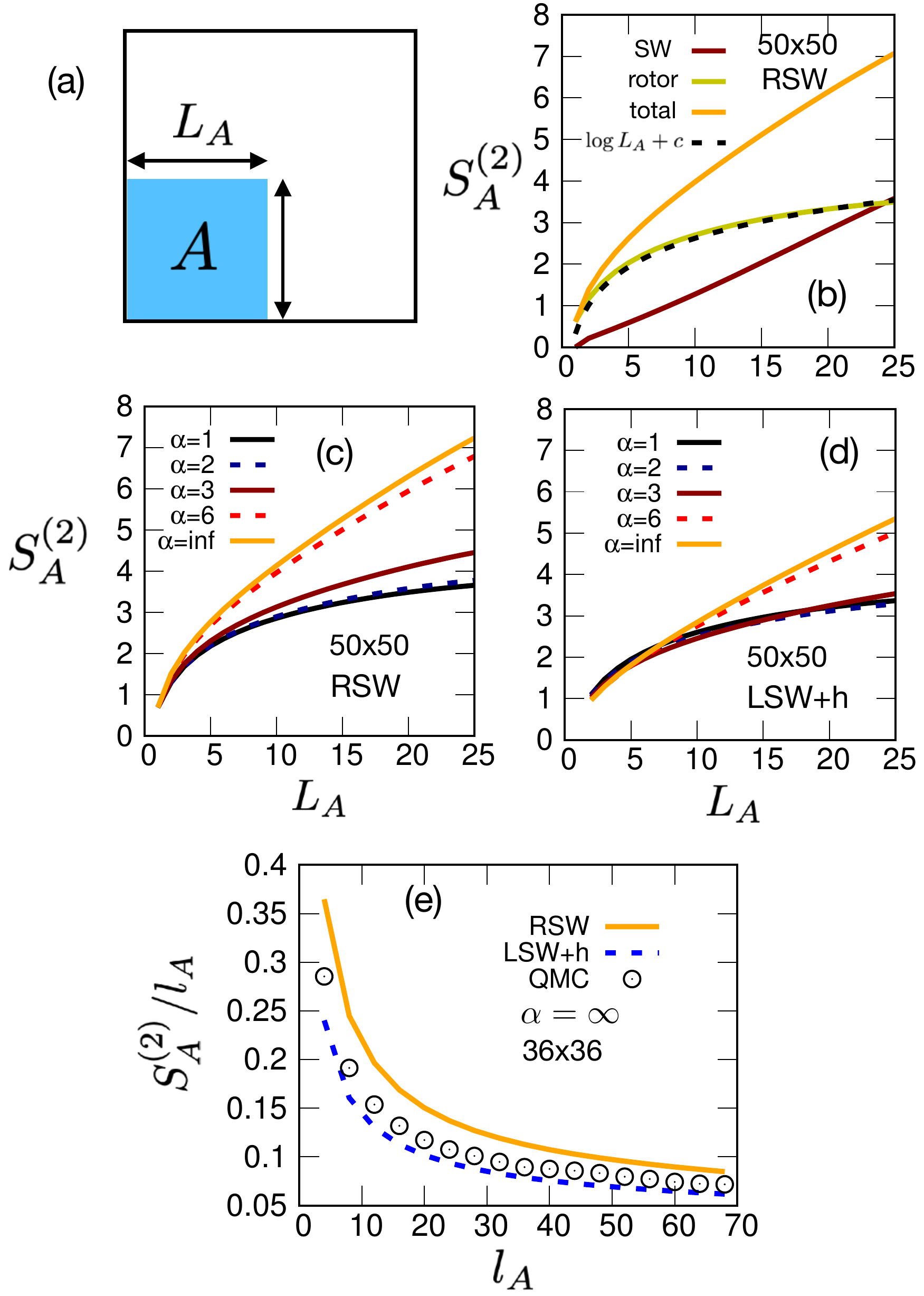} 
\caption{\emph{Ground-state entanglement entropy of the 2$d$ $\alpha-XX$ model.} (a) Geometry of the subsystem $A$ whose entropy is shown in the following panels; (b) Scaling of the entanglement R\'enyi entropy from RSW theory as a function of the linear dimension $L_A$ of the subsystem. The calculations have been performed for a system with nearest-neighbor interactions ($\alpha=\infty$) with linear dimension $L=50$. We show the SW contribution, the rotor one and their sum (giving the total entropy). The dashed line is a logarithmic fit; (c-d) Scaling of the entanglement R\'enyi entropy from RSW theory (c)  and LSW+h theory (d) for various values of $\alpha$; other parameters as in panel (b);  (e)   Scaling of the entanglement R\'enyi entropy with the subsystem perimeter length $l_A$ for a lattice with $L=36$. The predictions of RSW and LSW+h theory are compared with QMC results from Ref.~\cite{HumeniukR2012}.}
\label{f.GSentropy}
\end{center}
\end{figure}

\subsection{Ground-state energy and correlations of the 2d $\alpha$-XX model}

Fig.~\ref{f.equilibrium} shows the predictions of RSW theory for the ground state properties of the $\alpha$-XX model as a function of $\alpha$ and of the size $N=L\times L$ of the lattice with periodic boundary conditions.  
In particular we focus on the ground-state energy per spin $e_{\rm GS} = \langle H \rangle/(NJ)$; as well as on the correlation function at maximal distance
\begin{equation}
C_{L/2} = \frac{1}{2} \langle S^x_i S^x_{i+L/2}  + S^y_i S^y_{i+L/2} \rangle
\label{e.CL2}
\end{equation}
where $i+L/2$ indicates a site translated with respect to an arbitrary reference site $i$ by a distance $L/2$ along one of the two coordinate axes of the lattice. The net magnetization in the equilibrium state of a finite-size system is zero -- an exact result, imposed by construction within RSW theory as well as LSW+h theory. Hence, in the absence of a net magnetization, $C_{L/2}$ plays the role of a squared order parameter. 
The results of RSW theory are compared with those of LSW+h theory, as well as with QMC, offering the numerically exact reference. In particular QMC simulations are conducted at temperatures $T\sim {\cal O}(1/L)$ guaranteeing that thermal effects are eliminated from the finite-size results. 

Fig.~\ref{f.equilibrium}(a-b) shows that RSW theory and LSW+h theory provide nearly equivalent results on a sufficiently large lattice ($L=20$ in the figure in question) across a large spectrum of $\alpha$ values, from the long-range regime strictly defined ($\alpha <d=2$) to the short-range one $\alpha \gg d$); and that the results of the two theories are also in very good agreement with QMC. \footnote{We remark in passing that within LSW+h theory the two correlation functions $\langle S^x_i S^x_{i+L/2}\rangle$ and  $\langle S^y_i S^y_{i+L/2} \rangle$, building up the expression of $C_{L/2}$ in Eq.~\eqref{e.CL2}, are widely different, and they may even take unphysical values; it is only their average which, somewhat magically, reproduces correctly the QMC correlation function. On the other hand, within RSW theory the two correlation functions always take physical values, they become identical as $\alpha\to 0$, and they are closer to each other the smaller $\alpha$.}

In Fig.~\ref{f.equilibrium}(c-f) we show the finite-size scaling of the above-cited quantities for two physically relevant values of $\alpha$: $\alpha=3$ corresponding to dipolar interactions; and $\alpha=\infty$ corresponding to nearest-neighbor interactions. In both cases, we observe that ground-states energies predicted by the two theories are nearly identical. 
In particular we remark that both theories reproduce very well the ground-state energy for $\alpha=3$; but less so for for $\alpha=\infty$. Remarkably, both theories capture correctly the change in the scaling of the energy from decreasing with system size (for $\alpha=3$) to increasing (with $\alpha=\infty$), due to the change in the nature of the interactions from power-law decaying to short-ranged. On the other hand correlations predicted by RSW theory and LSW+h theory differ from each other on small system sizes, and they converge to the same value only for large sizes. Which theory best reproduces the QMC results seems to depend on the size range, but it is fair to say that RSW results are globally closer to the numerically exact ones. In particular RSW theory captures the non-monotonic size dependence of the long-distance correlations for  $\alpha=3$.

\subsection{Entanglement entropy}
\label{s.entropy}

We now move to a more advanced level of scrutiny of the predictions of RSW theory for the ground-state properties,  and we focus on the ground-state entanglement entropy for a bipartition. In particular we consider a subsystem $A$ which is a $L_A\times L_A$ square inside the $L\times L$ system, as shown in Fig.~\ref{f.GSentropy}(a). For the ground state of Hamiltonians breaking a continuous symmetry in the thermodynamic limit, the entanglement entropy is expected to scale as 
\begin{equation}
S^{(2)}_{A} = aL_A + b \log L_A + ...
\end{equation}
where $a$ is the coefficient of the dominant area-law term, while $b$ is the coefficient of the sub-dominant logarithmic term, which can be associated to the existence of $N_G$ gapless Goldstone modes. In the case of a $d$-dimensional system with Goldstone modes exhibiting a linear dispersion relation between frequency and wavevector, $\omega \sim k$, one has $b = N_G(d-1)/2$ \cite{GroverM2011}. If instead the Goldstone mode acquires a non-linear dispersion relation,  $\omega \sim k^z$, the $b$ coefficient is modified as  $b = N_G(d-z)/2$ \cite{Frerot2017PRB}. In the $\alpha$-XX model this occurs for $\alpha < d+2$: indeed $z = (\alpha-d)/2$ for $d < \alpha < d+2$, and $z=0$ for $\alpha \leq d$ \cite{Frerot2018PRL}. 

The RSW expression for the entanglement entropy of RSW theory, Eq.~\eqref{e.entRSW}, provides a natural decomposition into two differently scaling terms: as shown in Fig.~\ref{f.GSentropy}(b), the spin-wave term exhibits area-law scaling, while the rotor term exhibits a logarithmic scaling. 
Unfortunately the prefactor of the logarithmic scaling term is close to the subsystem entanglement entropy of a Dicke state $|J_{\rm tot}=NS,M=0\rangle$ -- even after the deformation of the rotor ground state by the application of a field (as in Eq.~\eqref{e.Hrotfield}) aimed at giving a net zero magnetization. Such an entropy is 
\begin{equation}
S^{(2)}_{A,\rm R} \approx \frac{1}{2} \log N_A = \frac{d}{2} \log L_A  
\end{equation}
where $N_A = L_A^d$ is the volume of the $A$ subsystem. This means that in the case of the $\alpha$-XX model with $N_G = 1$, the prefactor of the logarithmic term is $d/(d-z)$ times the one expected for $\alpha > d$, and it only coincides with the expected prefactor for $\alpha \leq d$ (namely when $z=d$), exhibiting again the fact that the predictions of RSW theory become exact in the small-$\alpha$ limit.  

The fact that the entanglement entropy of RSW theory overestimates systematically the logarithmic term for $d>\alpha$ is not surprising, in light of the discussion of Sec.~\ref{s.Hilbert}. It is a rather natural result of the extension of the Hilbert space with respect to that of the spin model of origin. As already discussed in Ref.~\cite{GroverM2011}, our result is a consequence of the sharp decoupling between the rotor variable and the spin-wave ones, which are postulated by RSW theory to give additive contributions to the entanglement entropy. The result $b = N_G(d-1)/2$ for the prefactor of the logarithmic term of Ref.~\cite{GroverM2011} descends instead from taking into account the interplay between the rotor (or ToS) spectrum of a subsystem, and the lowest-energy spin waves coupling the subsystem to its complement.
On the other hand, LSW+h theory is able to capture the correct prefactor of the logarithmic term, as shown in several recent works \cite{Songetal2011,Frerot2017PRB}. In spite of the fact that LSW+h theory fails to reproduce correctly the ToS spectrum -- as we will further show in Ref.~\ref{s.spectrum} -- it has the merit of treating the contributions of the zero mode and of the finite-wavevector modes to the entanglement entropy in a coupled manner. This appears to be sufficient to correctly recover the mechanism that leads to the appearance of the universal logarithmic contribution.    

Fig.~\ref{f.GSentropy}(c-d) shows the scaling of the second R\'enyi entropy as a function of subsystem size in the 2d $\alpha$-XX model for various values of $\alpha$, comparing the RSW predictions (Fig.~\ref{f.GSentropy}(b)) with those of LSW+h theory (Fig.~\ref{f.GSentropy}(c)). The two theories predict an increase of the area-law term upon growing $\alpha$, but within LSW+h theory the prefactor of the logarithmic term decreases with $\alpha$, leading to a non-monotonic $\alpha$-dependence of the subsystem entropies for sufficiently small subsystem sizes.  The discrepancy between the predictions of the two theories is once again to be largely attributed to the logarithmic term. This is particularly well seen in Fig.~\ref{f.GSentropy}(e), which shows the subsystem entanglement R\'enyi entropy as a function of the perimeter $l_A = 4(L_A-1)$ in the case of the XX model with nearest-neighbor interactions --  $\alpha=\infty$. When compared with QMC data from Ref.~\cite{HumeniukR2012}, one sees that RSW theory overestimates the numerically exact results, while LSW+h theory underestimates them. Yet the entropy per perimeter unit $S_A/l_A$ of both theories appears to converge to the exact result asymptotically, indicating that they can both reproduce the correct dominant area-law scaling term. 

\begin{figure*}[ht!]
\begin{center}
\includegraphics[width=0.95\textwidth]{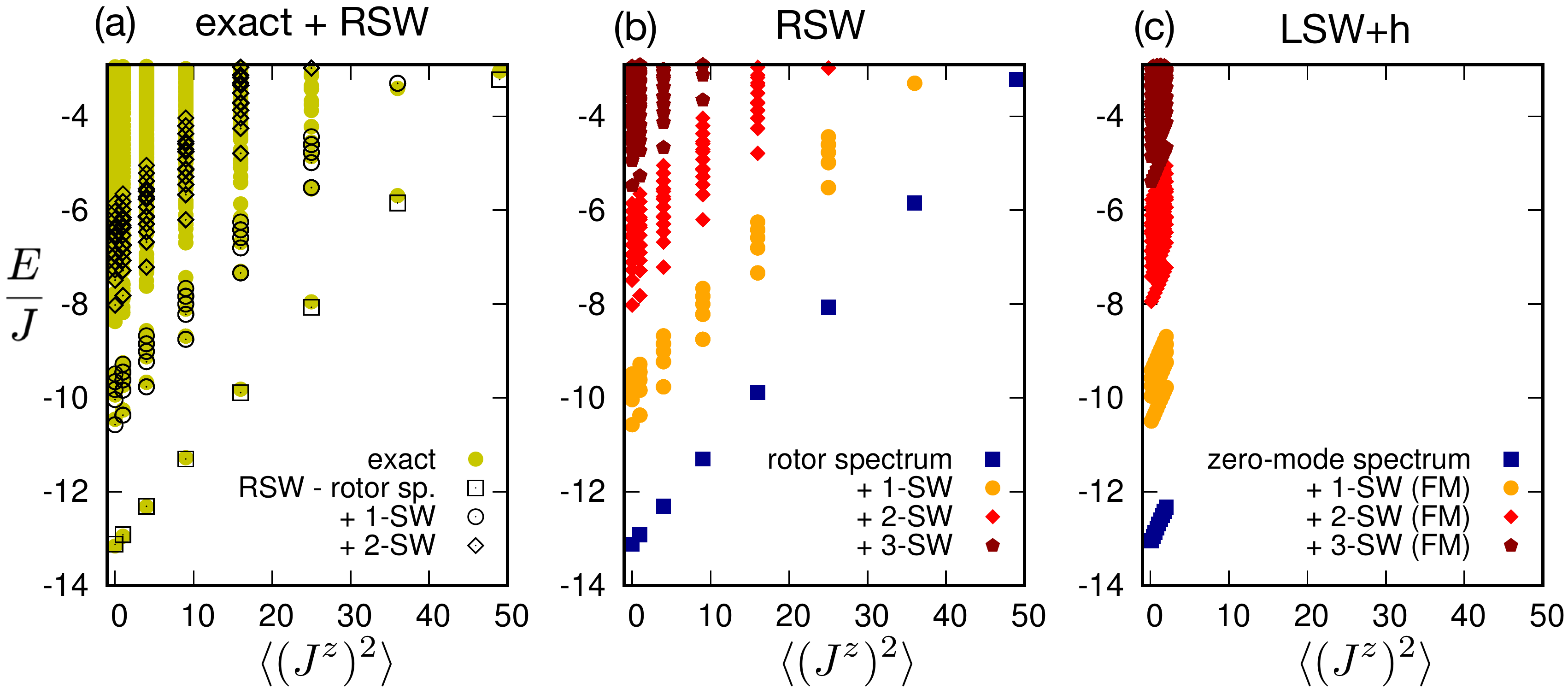}
\caption{\emph{Low-lying energy spectrum of the 2d XX model with dipolar interactions.} (a) Exact spectrum for a $L\times L$ lattice with $L=4$, plotted as a function of the $(J^z)^2$ quantum number. The RSW data of panel (b) (for the rotor spectrum, with addition of one-SW and two-SW excitations) are also shown for comparison; (b) RSW prediction for the spectrum, plotted as a function of $\langle (J^z)^2 \rangle$. The different symbols highlight the low-lying rotor spectrum; and the same spectrum with addition of one-SW energies, two-SW energies and three-SW energies. (c) LSW+h prediction for the spectrum. Same significance of the symbols as in (b), with the difference that the rotor spectrum is replaced by the harmonic zero-mode spectrum. }
\label{f.spectrum}
\end{center}
\end{figure*}




\section{Excitation spectrum}
\label{s.spectrum}

Next we analyze the low-energy spectrum of the two-dimensional $\alpha$-XX model with dipolar interactions, $\alpha=3$, comparing the exact result on a small ($L=4$) system with the predictions of RSW and LSW+h theory. Fig.~\ref{f.spectrum}(a) shows the energy levels of the exact spectrum plotted as a function $(J^z)^2$, which is a good quantum number given the U(1) symmetry of the problem. The spectrum resolved in terms of the $(J^z)^2$ quantum number clearly exhibits the existence of a branch of low-energy excitations with energy linear in  $(J^z)^2$ -- the Anderson's ToS, already cited in Sec.~\ref{s.intro}, composed of the ground states of the Hamiltonian in each $J^z$ sector. The energy spectrum of the ToS is, to a very good approximation, given by  
\begin{equation}
E_{\rm ToS}(J^z) \approx E_{\rm gs,ex} + \frac{(J^z)^2}{2I_{\rm ToS}}~
\end{equation}
where $E_{\rm gs,ex}$ is the exact ground-state energy and $I_{\rm ToS}$ should be thought of as the effective moment of inertia of a planar-rotor variable having the same spectrum as that of the ToS. 
A further, striking feature of the spectrum is the fact that the same, nearly linear dependence on $(J^z)^2$ can be found for further groups of higher-energy states, forming approximate towers which are just shifted by a constant with respect to the low-energy ToS. 

This picture is clearly compatible with the one offered by RSW theory, Eq.~\eqref{e.RSWspectrum}, for which the energy shifts between successive towers of states is given by sums of spin-wave frequencies. As such it is therefore strongly suggestive of an (approximate) additive structure of the spectrum, resulting from the sum of a rotor contribution and a SW one. The quantitative correspondence between the low-energy part of the RSW spectrum and the exact one is indeed shown in Fig.~\ref{f.spectrum}(b); while  the same RSW spectrum is repeated for clarity in Fig.~\ref{f.spectrum}(b). The RSW energies are plotted as a function of $\langle (J^z)^2 \rangle \approx \langle (K^z)^2 \rangle$, where $\langle... \rangle$ is the average on the excited state. This average replaces the quantum number $(J^z)^2$ since we slightly broke the U(1) symmetry of the rotor Hamiltonian by applying a transverse field term which compensates the SW magnetization in the ground state, as discussed in Sec.~\ref{s.modification}. As it is apparent in the figure, the values of $\langle (J^z)^2 \rangle$ remain very close to squares of integer numbers \footnote{In principle the construction leading to a vanishing total magnetization should be done for each excited state, resulting in an energy-dependent transverse field -- yet for simplicity we plot the spectrum obtained using a field which cancels the $\langle J^x \rangle$ magnetization only in the ground state, which implies that the excited states have in fact a net magnetization. This choice is coherent with the fact that, when studying the dynamics, we do not apply any field, so that the excitation spectrum shown in Fig.~\ref{f.spectrum} is essentially the one that will manifest itself in the time evolution of observables.}.
As expected from the rotor contribution, the energy levels depend linearly on $\langle (J^z)^2\rangle $, with a slope given by $1/(2I)$.

 As shown in Fig.~\ref{f.spectrum}(a), the ``bare" moment of inertia of the rotor $I$ predicted by RSW theory, Eq.~\eqref{e.1_2Itilde}, does not perfectly coincide with the one emerging from the ToS of the exact spectrum $I_{\rm ToS}$. In fact we observe in general that $I > I_{\rm ToS}$. For the $N=16$ data in Fig.~\ref{f.spectrum}, $I =  2.47 J^{-1}$ and $I_{\rm ToS} =  2.42 J^{-1}$, hence the difference is rather small in this case. But, as shown by us in Ref.~\cite{Comparin2022PRA}, the discrepancy between the two moments of inertia grows with $\alpha$; \emph{e.g.} for  $\alpha = \infty$, $I =  3.75 J^{-1}$ while $I_{\rm ToS} =  3.45 J^{-1}$.  This can be readily interpreted as a renormalization effect due to the residual interactions of the rotor with the finite-momentum spin waves, which we neglect within RSW theory. Taking into account this renormalization in a scalable way is indeed possible, as further discussed by us in Ref.~\cite{Roscildeetal2023}.  

In Fig.~\ref{f.spectrum}(b) we group the excited states into bands of ``towers", representing the low-lying (or strictly defined) ToS shifted by one SW excitation, two SW excitations, three SW excitations, etc.  The comparison with Fig.~\ref{f.spectrum}(a) shows that one can clearly identify the one-SW-excitation branch in the exact spectrum, separated by a gap from the branches with a higher number of SW excitations (at least for sufficiently small $(J^z)^2$). The distinction between branches with more than two spin-wave excitations is instead lost in the exact spectrum. 

Finally, Fig.~\ref{f.spectrum}(c) shows the prediction of LSW+h theory for the spectrum. Within this theory the lowest-energy excitations are obtained by populating the zero mode at energy $\epsilon_{\bm q=0} \sim N^{-1}$ -- see Sec.~\ref{s.LSWh} -- and have energies $E_0 + n \epsilon_{\bm q=0}$ with $n=1,2,...$, 
where $E_0$ is the LSW+h ground-state energy given by Eq.~\eqref{e.E0SW}.
The higher-energy states are obtained by adding finite-momentum spin waves to each sector with $n$ zero-mode excitations, resulting in the group of states with one SW, two SWs, etc. -- as indicated in Fig.~\ref{f.spectrum}(c) \footnote{Similarly to our treatment of the RSW spectrum, we choose a field $h$ leading to a vanishing magnetization (see Sec.~\ref{s.LSWh}) only in the ground state, and we show the rest of the spectrum as calculated at the same fixed field. In principle one should find a field $h$ leading to a vanishing magnetization for each energy level, but for simplicity we did not pursue this calculation. In the same spirit as for the remark made above on RSW theory, the spectrum without any $h$ field is in fact the one relevant for the dynamics.}. As it is clear from the picture, the SW ``bands" are correctly reproduced -- as one should expect, since the spin-wave spectrum at finite momentum is essentially the same as in RSW theory. But the ToS structure is completely incorrect, given that the zero mode is represented as a (linear) harmonic oscillator in LSW+h theory, and as such it cannot provide an accurate approximation for the (non-linear) rotor spectrum of the actual low-lying ToS, if not for the very first excitation. In particular the values of $\langle (J^z)^2 \rangle$
for the excited states are completely wrong, because of the incorrect treatment of this zero mode beyond the description of the ground state. 

 The ability of RSW theory to correctly reproduce the low-lying excitation spectrum is the key aspect behind its success in reproducing the low-energy non-equilibrium dynamics initialized in the CSS, as we will discuss in the next section.

\begin{figure*}[ht!]
\begin{center}
\includegraphics[width=0.95 \textwidth]{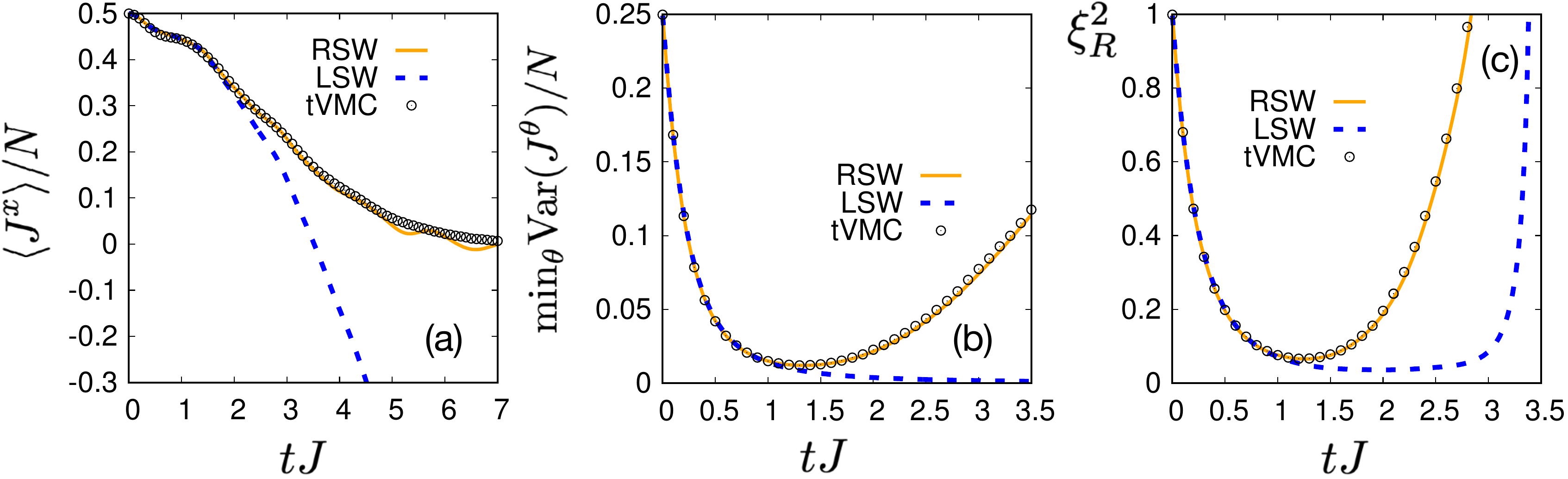}
\caption{\emph{Collective-spin dynamics after a quench in the 2d dipolar XX model. } (a) Relaxation of the average collective spin for a square lattice with $L=10$. The panel compares the prediction of RSW and LSW theory with the results of tVMC; (b) Minimum variance of the collective-spin components transverse to the average orientation; (c) Spin squeezing parameter. For both panels (b-c), same lattice geometry and significance of symbols as in (a).}
\label{f.dynamics}
\end{center}
\end{figure*}

\section{Quench dynamics}
\label{s.quench}

In this section we test the physical picture underpinning RSW theory -- the approximate separation between a non-linear zero-mode degree of freedom and linearized finite-momentum spin waves --  in the case of non-equilibrium dynamics.
 Given that RSW theory describes the low-energy properties of the system, we choose to examine a very significant instance of low-energy quench dynamics, namely the dynamics of the dipolar XX model ($\alpha=3$) initialized in the coherent spin state (CSS), representing the vacuum of Holstein-Primakoff bosons (Sec.~\ref{s.spinboson}). A detailed study of this dynamics has been presented by us in the companion paper, Ref.~\cite{Roscildeetal2023}. 
 The goal of this section is to compare RSW predictions with the results of time-dependent variational Monte Carlo (tVMC), which offers a very accurate solution to the dynamics of the system, as shown by us in Ref.~\cite{Comparin2022PRL}; and to contrast them with the predictions of standard LSW theory, which amounts to linearizing the zero mode. 


\subsection{Non-equilibrium RSW theory and OAT dynamics}

Within RSW theory, the dynamics of the XXZ model is simply represented as the independent non-linear dynamics of the rotor, described by the one-axis-twisting Hamiltonian \cite{Kitagawa1993PRA} of Eq.~\eqref{e.HZM}; and that of linear spin waves at finite momentum. Both dynamics can be calculated efficiently: the rotor variable lives in a $(2NS+1)$-dimensional Hilbert space; while studying the dynamics of linear spin waves amounts to solving ${\cal O}(N/2)$ pairs of coupled differential equations for the regular and anomalous correlators $G_{\bm q} = \langle b^\dagger_{\bm q} b_{\bm q} \rangle$ and $F_{\bm q} = \langle b_{\bm q} b_{-\bm q} \rangle$  associated with two HP-boson modes at opposite momenta ${\bm q}$ and $-{\bm q}$. The equations are given in App. \ref{a.SW}, along with their analytical solution via Bogolyubov diagonalization at finite momentum \cite{Frerot2018PRL}. 

 For RSW theory to be valid, the density of finite-momentum bosons $N_{\rm FM}/N$ must remain very low along the dynamics, so as to justify the assumption of decoupling between spin waves and rotor. On the other hand the rotor dynamics is completely arbitrary, since its nonlinearities are fully accounted for. If the rotor/spin-wave decomposition of all observables, as defined in Sec.~\ref{s.observables}, is dominated by the rotor contribution, then the dynamics of the entire system is akin to that of the one-axis-twisting model \cite{Kitagawa1993PRA}, which features in particular the appearance of scalable spin squeezing at short times, signaled by the spin-squeezing parameter \cite{Wineland1994PRA} 
 \begin{equation}
 \xi^2_R = \frac{N \min_{\theta} {\rm Var}(J^\theta)}{\langle J^x \rangle^2}
 \label{e.xi2R}
 \end{equation}    
 where $J^\theta = \cos\theta J^y + \sin\theta J^z$, so that   
 \begin{align}
 {\rm Var}(J^\theta) = & \cos^2\theta ~{\rm Var}(J^y) +  \sin^2\theta ~{\rm Var}(J^z) \nonumber \\
 & + 2 \sin\theta \cos\theta ~{\rm Cov}(J^y,J^z)~.
 \end{align}
 Spin squeezing is associated with the condition $\xi_R^2 < 1$, which witnesses the presence of entanglement \cite{Sorensen2001}; in particular the OAT dynamics features minimal squeezing parameter attained at times $t_{\rm sq} \sim N^{1/3}$, and scaling as $(\xi_R^2)_{\rm min} \sim N^{-2/3}$ (for very large $N$) \cite{Kitagawa1993PRA}. 
 
 According to the discussion of Sec.~\ref{s.observables}, within RSW theory the spin-squeezing parameter is simply given by 
 \begin{equation}
 \xi^2_R \approx \frac{N \min_{\theta} {\rm Var}(K^\theta)_{\rm R}}{(\langle K^x \rangle_{\rm R}-N_{\rm FM})^2}
 \end{equation}  
 namely it differs from the spin squeezing parameter of the OAT model by the fact that the average spin is renormalized by the spin-wave contribution.  
 
\subsection{Non-equilibrium LSW theory and squeezing dynamics}

Non-equilibrium LSW theory amounts essentially to treating the rotor variable in a linearized zero-momentum HP boson, at the same level as the finite-momentum HP bosons. 
When dealing with (short-time) non-equilibrium dynamics, one can ignore the complication associated with the Bogolyubov diagonalization of the zero-momentum bosons (see Sec.~\ref{s.LSWh}), and simply extend the linearized equations of motion for the HP bosons (App. \ref{a.SW}) to the ${\bm q}=0$ one. 
A similar approach has been used in a variety of recent studies, and it is successful for systems that do not possess the zero-mode pathology descending from U(1) symmetry \cite{MenuR2018,MenuR2020,Cevolanietal2019,Lerose2019PRB,Lerose2020PRR}. For systems with U(1) symmetry, gapping out the zero mode as in LSW+h theory is not an option away from equilibrium, because the application of a field is only justified when requesting the average collective spin to vanish.The latter average is instead maximal at $t=0$ in the particular quench scheme that we are considering. Hence the zero-mode pathology necessarily limits the validity of LSW theory to short times when dealing with finite-size systems.  

As already shown in Eq.~\eqref{e.H2q}, the Hamiltonian governing the linearized dynamics of the ${\bm q}=0$ boson has the form
\begin{equation}
{\cal H}_0 = A_0 b_0^\dagger b_0 + \frac{B_0}{2} \left [ (b_0^\dagger)^2 + (b_0)^2 \right ]   ~.
\label{e.H0}
\end{equation}
{
which is a squeezing Hamiltonian as known in quantum optics \cite{WallsMilburn}. Indeed, as it can be verified from their expressions in Sec.~\ref{s.SWtheory}, one has that $A_0 = -B_0 =  SJ_0(1-\Delta)/2 = NS/2\tilde{I}$, so that, introducing the (dimensionless) $P$ quadrature of the bosonic field, $P = (b_0-b_0^\dagger)/(i\sqrt{2})$, one has that ${\cal H}_0 = NS/(2\tilde{I}) ~P^2 + {\rm const.} \approx (J^z)^2/2\tilde{I} + {\rm const.}$, where we have used the linearized HP transformation $J^z \approx \sqrt{2SN} (b_0-b_0^\dagger)/(2i) = \sqrt{NS} P$. The ${\cal H}_0 \sim P^2$ Hamiltonian evolves the initial vacuum state -- which in the position eigenbasis is a Gaussian wavepacket corresponding to the ground state of a harmonic oscillator -- via an indefinite free expansion. The latter leads to squeezing of a quadrature of the field \cite{Garbeetal2022}, intermediate between $P$ and $X =  (b_0+b_0^\dagger)/\sqrt{2} \approx J^y/\sqrt{NS}$, corresponding to the squeezing of one collective-spin component in the $yz$ plane. This is expected as ${\cal H}_0$ is the quadratic approximation to the OAT Hamiltonian. Nonetheless, at the same time the fluctuations of $X$ grow unboundedly under free evolution, which implies that the number of zero-momentum bosons $N_0 = \langle b_0^\dagger b_0 \rangle =  \langle X^2 \rangle + \langle P^2 \rangle - 1/2$ also grows without limits, driving a runaway of the average spin $\langle J^x \rangle = NS - N_0 - N_{\rm FM}$. Here $N_{\rm FM}$ is the number of finite-momentum bosons, as defined in Sec.~\ref{s.observables}.  This proliferation of HP bosons at zero momentum exposes again the zero-mode pathology of LSW in the dynamics.  }

\subsection{Results}

Fig.~\ref{f.dynamics} offers a comparison between the results of RSW theory, LSW theory and tVMC based on a pair-product wavefunction \cite{Thibaut2019PRB, Comparin2022PRA, Comparin2022PRL} for the quench dynamics of the $S=1/2$ 2d dipolar XX model initialized in the CSS. We focus in particular on a square lattice comprising $N=100$ sites.  
In particular, Fig.~\ref{f.dynamics}(a) shows the depolarization dynamics of the average collective spin, which is very well reproduced by RSW theory, because the rotor contribution to it obeys the non-linear dynamics of the OAT Hamiltonian. On the other hand, the dynamics of zero-momentum bosons is governed by the linear Hamiltonian of Eq.~\eqref{e.H0} within LSW theory, which leads to an unbounded proliferation of such bosons, and therefore to a much too fast depolarization of the collective spin. 

Concomitantly, one of the transverse components of the collective spin develops a squeezed uncertainty below the shot-noise limit ${\rm Var}(J^\theta)/N  = S/2 = 1/4$ of the initial state. As shown in Fig.~\ref{f.dynamics}(b), this behavior is well reproduced by both RSW and LSW theory at short times. But at longer times the two theories depart from each other. RSW theory predicts that $\min_{\theta} {\rm Var}(J^\theta)$ obeys the same dynamics as that of the squeezed spin component of the OAT model -- and this prediction is confirmed by tVMC. { On the other hand the minimum variance within LSW theory continues decreasing indefinitely, because squeezing occurs here on the infinite quadrature plane, as opposed to spin squeezing, which occurs on the finite Bloch sphere.} 

{ The squeezing parameter $\xi_R^2$ of Eq.~\eqref{e.xi2R} is then built out of the ratio between the two quantities considered so far. While the RSW results are in excellent agreement with the tVMC, we see that the LSW results only reproduce the squeezing at the beginning of the evolution and until the minimum of the $\xi_R^2$ parameter. Successively, the LSW prediction for  $\xi_R^2$ continues decreasing below the actual minimum, driven by the unbounded decrease of the minimum variance; until the vanishing of the average spin leads to a sharp, unbounded increase in $\xi_R^2$. 

These results show that the LSW dynamics ceases to be quantitative after a short time. On the other hand RSW dynamics remains quantitative up to macroscopic times $\sim {\cal O}(N)$, as discussed in details in our companion paper Ref.~\cite{Roscildeetal2023}.}

\section{Conclusions}
\label{s.conclusions}

In this work we have introduced a new approach - rotor/spin-wave (RSW) theory  -- to treat the finite-size equilibrium and non-equilibrium behavior of quantum spin systems with U(1) symmetry. Similarly to conventional spin-wave theory, our approach maps spin deviations with respect to a classical reference state onto a gas of bosonic quasiparticles. Yet, unlike spin-wave theory, it only assumes that this gas is dilute at finite momenta. Zero-momentum bosons are instead treated with all their non-linearities within RSW theory; and they are shown to reconstruct a quantum-rotor degree of freedom, which in principle can take arbitrarily non-classical states. The spectrum of the quantum rotor reconstructs the Anderson tower of states of excitations, characteristic of finite-size systems which spontaneously break a U(1) symmetry in the thermodynamic limit. Hence our theory can account simultaneously for linear spin-wave excitations at finite momentum as well as for the non-linear ToS excitations, fully reconstructing the low-energy spectrum.  

RSW theory reproduces very well the ground-state properties for U(1)-symmetric systems displaying long-range order, in a way which is similar (and on some accounts superior) to conventional spin-wave theory modified for finite-size systems. But it proves to be far superior to spin-wave theory in describing the low-energy spectrum, as well as the non-equilibrium dynamics. In particular it is able to correctly describe the highly non-linear quantum dynamics by which a finite-size system effectively restores its U(1) symmetry broken in the initial state, by relaxing towards an unpolarized -- i.e. nonclassical -- state. In this work we have applied the RSW theory to U(1)-symmetric $S=1/2$ spin systems. Yet the theory is readily generalizable to arbitrary spin lengths, and it can be extended to higher symmetries (\emph{e.g.} SU(2)), as we will discuss in a future work. 

The physical picture emerging from RSW theory is that of an effective separation of variables between zero-momentum degrees of freedom (representing the quantum-rotor variable) and finite-momentum ones, corresponding to linear spin waves. This picture appears to be valid when the finite-momentum spin waves form a very dilute gas, but it is expected to become less and less accurate for higher densities of such spin waves, corresponding to increasingly strong fluctuations (classical or quantum) of the spins at finite momentum. Restricting to pure states, the strength of these finite-momentum (quantum) fluctuations can be controlled by the connectivity of the lattice, i.e. the range of the interactions and/or its dimensionality. RSW theory becomes asymptotically exact in the limit of infinite-range interactions / infinite dimensions, while it is less accurate for short-range interactions. Systematic improvement on RSW theory can be achieved by taking into account explicitly the coupling between the rotor variable and the finite-momentum spin waves. Pushing forward this program completely would be as hard as exactly diagonalizing the system; but one can envision to take into account the coupling of the rotor with only a subset of finite-momentum modes, thereby increasing progressively the computational complexity of the approach.   

Another challenging aspect for our approach is to generalize it beyond the case of translationally invariant systems treated here. Breaking of translational invariance introduces a coupling between the rotor and the finite-momentum spin waves at quadratic order in the bosonic operators, which cannot be reasonably neglected, unless it is a boundary term in a large system. Taking into account explicitly the coupling between rotor and spin waves (or a subset thereof) is therefore necessary to extend the theory to treat \emph{e.g.} disordered systems. 

 The ability of RSW theory (and of its potential future extensions) to deal with highly non-classical states of spin systems with very light computational resources can offer a very valuable guidance for experimental quantum simulation of dynamics of lattice quantum spin systems -- based e.g. on arrays of Rydberg atoms, of trapped ions, of neutral atoms in optical lattices, or of superconducting qubits, to cite some relevant examples. Its physical content is very transparent, and it allows one to reach very large system sizes, comprising up to thousands of spins, in a very short computational time. It offers therefore a valuable alternative to more sophisticated and computationally demanding approaches for the equilibrium and non-equilibrium properties of quantum simulators, such as semi-classical methods 
\cite{Schachenmayer2015PRX} or variational ones \cite{Carleo2014PRA,Paeckeletal2019,Comparin2022PRA}. 

\begin{acknowledgements}
Useful discussions and collaborations on related subjects with Y. Trifa and J. Br\'ehier are gratefully acknowledged. This project is supported by ANR (EELS project), QuantERA (MAQS project) and PEPR-Q (QubitAF project). All numerical calculations have been performed on the PSMN cluster at ENS de Lyon.   
\end{acknowledgements}

\medskip
\medskip
\medskip

\appendix

\section{Correlation functions within RSW theory}
\label{a.correlations}

This prescription of Eq.~\eqref{e.correlations} for the correlation functions within RSW theory leads to the formulas 
\begin{align}
C^{xx}_{ij} & \approx \langle [(S^x_i)^2]_{\rm ZM}\rangle_{\rm R}~\delta_{ij}+   \\
& + \frac{ \langle (K^x)^2 \rangle_{\rm R} - \sum_l \langle [(S_l^x)^2]_{\rm ZM}\rangle_{\rm R}}{N(N-1)} ~(1-\delta_{ij})   \nonumber \\
 & - \frac{2S}{N} \sum_{\bm q \neq 0} \langle b^\dagger_{\bm q}  b_{\bm q} \rangle_{\rm SW} \nonumber 
\end{align}
\begin{align}
C^{yy}_{ij} & \approx   \langle [(S^y_i)^2]_{\rm ZM}\rangle_{\rm R}~\delta_{ij}+   \\
& +  \frac{ \langle (K^y)^2 \rangle_{\rm R} - \sum_l \langle [(S_l^y)^2]_{\rm ZM}\rangle_{\rm R}}{N(N-1)}   ~(1-\delta_{ij})  \nonumber  \\
& +  \frac{S}{2N} \sum_{\bm q\neq 0} e^{i\bm q \cdot (\bm r_i - \bm r_j)} \Big ( \langle b^\dagger_{\bm q}  b_{\bm q} \rangle_{\rm SW}  
+ \langle b^\dagger_{\bm q}  b^\dagger_{-\bm q} \rangle_{\rm SW} + {\rm c.c.} \Big )  \nonumber  
\end{align}
\begin{align}
C^{zz}_{ij} & \approx     \langle [(S^y_i)^2]_{\rm ZM}\rangle_{\rm R}~\delta_{ij}   \\ 
& + \frac{ \langle (K^z)^2 \rangle_{\rm R} - \sum_l \langle [(S_l^z)^2]_{\rm ZM}\rangle_{\rm R}}{N(N-1)} ~(1-\delta_{ij}) \nonumber \\
& +  \frac{S}{2N} \sum_{\bm q\neq 0} e^{i\bm q \cdot (\bm r_i - \bm r_j)} \Big ( \langle b^\dagger_{\bm q}  b_{\bm q} \rangle_{\rm SW}  
- \langle b^\dagger_{\bm q}  b^\dagger_{-\bm q} \rangle_{\rm SW} + {\rm c.c.} \Big )  \nonumber  
\end{align}
\begin{align}
C^{xy}_{ij} & \approx \frac{1}{2} \langle [ \{ S_i^x, S_i^y\}]_{\rm ZM}\rangle_{\rm R}~\delta_{ij}  \\
& +    \frac{\langle  \{ K^x, K^y\}\rangle_{\rm R} - \sum_i \langle [\{S_i^x ,S_i^y \}]_{\rm ZM}\rangle_{\rm R} }{2N(N-1)} ~(1-\delta_{ij}) \nonumber  
\end{align}
\begin{align}
C^{xz}_{ij} & \approx \frac{1}{2}  \langle [ \{ S_i^x, S_i^z\}]_{\rm ZM}\rangle_{\rm R}~\delta_{ij} \\
& +    \frac{\langle  \{ K^x, K^z\}\rangle_{\rm R} - \sum_i \langle [\{S_i^x ,S_i^z \}]_{\rm ZM}\rangle_{\rm R} }{2N(N-1)} ~(1-\delta_{ij})   \nonumber  
\end{align}
\begin{align}
C^{yz}_{ij} & \approx    \frac{1}{2}  \langle [ \{ S_i^y, S_i^z\}]_{\rm ZM}\rangle_{\rm R}~\delta_{ij} \\
&+ \frac{  \langle  \{ K^y, K^z\}\rangle_{\rm R} - \sum_i \langle [\{S_i^y ,S_i^z \}]_{\rm ZM}\rangle_{\rm R}  }{2N(N-1)} ~(1-\delta_{ij})  \nonumber \\ 
& + \frac{S}{2iN} \sum_{\bm q\neq 0} e^{i\bm q \cdot (\bm r_i - \bm r_j)} \Big ( \langle b_{\bm q}  b_{-\bm q} \rangle_{\rm SW} 
 -  \langle b^\dagger_{\bm q}  b^\dagger_{-\bm q} \rangle_{\rm SW} \Big ) ~.  \nonumber 
\end{align} 
It is immediate to verify that the above expressions capture correctly the CSS expectation values,  i.e. $C^{xx}_{ij} = S^2$, $C^{yy}_{ij} = C^{zz}_{ij} = (S/2) \delta_{ij}$, and
$C^{xy}_{ij} = C^{xz}_{ij}  = C^{yz}_{ij} = 0$.

\section{Equations of motion for the Holstein-Primakoff bosons within LSW}
\label{a.SW}

Here we provide the equations of motion for the correlators $G_{\bm q} = \langle b^\dagger_{\bm q} b_{\bm q} \rangle$ and $F_{\bm q} = \langle b_{\bm q} b_{-\bm q} \rangle$ under the dynamics governed by the linear Hamiltonian Eq.~\eqref{e.Hsw}. These correlators describe completely the Gaussian state of linearized HP bosons. The equations read:
\begin{eqnarray}
\frac{dG_{\bm q}}{dt} & = & -2 ~B_{\bm q} ~{\rm Im}(F_{\bm q})  \\
\frac{dF_{\bm q}}{dt} & = & -i  \left [ 2 A_{\bm q} F_{\bm q} + B_{\bm q} (1 + G_{\bm q} + G_{-\bm q} ) \right ] \nonumber~.
\label{e.TDLSW}
\end{eqnarray}
{
For finite momenta, these equations can be solved via the Bogolyubov transformation of Sec.~\ref{s.SWtheory}, to give 
\begin{align}
G_{\bm q}(t) &  = 2u_{\bm q}^2 v_{\bm q}^2 [ 1-\cos(2\epsilon_{\bm q}t)] \nonumber \\
F_{\bm q}(t) & = u_{\bm q} v_{\bm q} \left ( u_{\bm q}^2 e^{-2i\epsilon_{\bm q} t} + v_{\bm q}^2 e^{2i\epsilon_{\bm q} t} - 2v_{\bm q}^2 -1 \right )~.
\end{align}
For zero momentum the Bogolyubov transformation becomes singular; yet the equations for $F_0$ and $B_0$ reduce to those of a single bosonic mode subject to the squeezing Hamiltonian Eq.~\eqref{e.H0}. As mentioned in the main text, the solution to the dynamics is the same as that for the free expansion of the minimal-uncertainty Gaussian wavepacket, which is the vacuum state when expressed in the basis of the position operator $X = (b_0 + b_0^\dagger)/\sqrt{2}$. }

\bibliography{SWrotor.bib}

\begin{thebibliography}{49}
\expandafter\ifx\csname natexlab\endcsname\relax\def\natexlab#1{#1}\fi
\expandafter\ifx\csname bibnamefont\endcsname\relax
  \def\bibnamefont#1{#1}\fi
\expandafter\ifx\csname bibfnamefont\endcsname\relax
  \def\bibfnamefont#1{#1}\fi
\expandafter\ifx\csname citenamefont\endcsname\relax
  \def\citenamefont#1{#1}\fi
\expandafter\ifx\csname url\endcsname\relax
  \def\url#1{\texttt{#1}}\fi
\expandafter\ifx\csname urlprefix\endcsname\relax\def\urlprefix{URL }\fi
\providecommand{\bibinfo}[2]{#2}
\providecommand{\eprint}[2][]{\url{#2}}

\bibitem[{\citenamefont{Auerbach}(2006)}]{Auerbachbook}
\bibinfo{author}{\bibfnamefont{A.}~\bibnamefont{Auerbach}},
  \emph{\bibinfo{title}{Interacting Electrons and Quantum Magnetism}}
  (\bibinfo{publisher}{Springer}, \bibinfo{year}{2006}).

\bibitem[{\citenamefont{Blundell}(2001)}]{Blundellbook}
\bibinfo{author}{\bibfnamefont{S.}~\bibnamefont{Blundell}},
  \emph{\bibinfo{title}{Magnetism in Condensed Matter}}
  (\bibinfo{publisher}{OUP Oxford}, \bibinfo{year}{2001}).

\bibitem[{\citenamefont{Georgescu et~al.}(2014)\citenamefont{Georgescu, Ashhab,
  and Nori}}]{Georgescuetal2014}
\bibinfo{author}{\bibfnamefont{I.~M.} \bibnamefont{Georgescu}},
  \bibinfo{author}{\bibfnamefont{S.}~\bibnamefont{Ashhab}}, \bibnamefont{and}
  \bibinfo{author}{\bibfnamefont{F.}~\bibnamefont{Nori}},
  \bibinfo{journal}{Rev. Mod. Phys.} \textbf{\bibinfo{volume}{86}},
  \bibinfo{pages}{153} (\bibinfo{year}{2014}),
  \urlprefix\url{https://link.aps.org/doi/10.1103/RevModPhys.86.153}.

\bibitem[{\citenamefont{Mazurenko et~al.}(2017)\citenamefont{Mazurenko, Chiu,
  Ji, Parsons, Kan{\'{a}}sz-Nagy, Schmidt, Grusdt, Demler, Greif, and
  Greiner}}]{Mazurenko2017}
\bibinfo{author}{\bibfnamefont{A.}~\bibnamefont{Mazurenko}},
  \bibinfo{author}{\bibfnamefont{C.~S.} \bibnamefont{Chiu}},
  \bibinfo{author}{\bibfnamefont{G.}~\bibnamefont{Ji}},
  \bibinfo{author}{\bibfnamefont{M.~F.} \bibnamefont{Parsons}},
  \bibinfo{author}{\bibfnamefont{M.}~\bibnamefont{Kan{\'{a}}sz-Nagy}},
  \bibinfo{author}{\bibfnamefont{R.}~\bibnamefont{Schmidt}},
  \bibinfo{author}{\bibfnamefont{F.}~\bibnamefont{Grusdt}},
  \bibinfo{author}{\bibfnamefont{E.}~\bibnamefont{Demler}},
  \bibinfo{author}{\bibfnamefont{D.}~\bibnamefont{Greif}}, \bibnamefont{and}
  \bibinfo{author}{\bibfnamefont{M.}~\bibnamefont{Greiner}},
  \bibinfo{journal}{Nature} \textbf{\bibinfo{volume}{545}},
  \bibinfo{pages}{462} (\bibinfo{year}{2017}),
  \urlprefix\url{https://doi.org/10.1038/nature22362}.

\bibitem[{\citenamefont{Lepoutre et~al.}(2019)\citenamefont{Lepoutre,
  Schachenmayer, Gabardos, Zhu, Naylor, Mar{\'e}chal, Gorceix, Rey, Vernac, and
  Laburthe-Tolra}}]{Lepoutreetal2019}
\bibinfo{author}{\bibfnamefont{S.}~\bibnamefont{Lepoutre}},
  \bibinfo{author}{\bibfnamefont{J.}~\bibnamefont{Schachenmayer}},
  \bibinfo{author}{\bibfnamefont{L.}~\bibnamefont{Gabardos}},
  \bibinfo{author}{\bibfnamefont{B.}~\bibnamefont{Zhu}},
  \bibinfo{author}{\bibfnamefont{B.}~\bibnamefont{Naylor}},
  \bibinfo{author}{\bibfnamefont{E.}~\bibnamefont{Mar{\'e}chal}},
  \bibinfo{author}{\bibfnamefont{O.}~\bibnamefont{Gorceix}},
  \bibinfo{author}{\bibfnamefont{A.~M.} \bibnamefont{Rey}},
  \bibinfo{author}{\bibfnamefont{L.}~\bibnamefont{Vernac}}, \bibnamefont{and}
  \bibinfo{author}{\bibfnamefont{B.}~\bibnamefont{Laburthe-Tolra}},
  \bibinfo{journal}{Nature Communications} \textbf{\bibinfo{volume}{10}},
  \bibinfo{pages}{1714} (\bibinfo{year}{2019}), ISSN \bibinfo{issn}{2041-1723},
  \urlprefix\url{https://doi.org/10.1038/s41467-019-09699-5}.

\bibitem[{\citenamefont{Jepsen et~al.}(2020)\citenamefont{Jepsen, Amato-Grill,
  Dimitrova, Ho, Demler, and Ketterle}}]{Jepsen2020}
\bibinfo{author}{\bibfnamefont{P.~N.} \bibnamefont{Jepsen}},
  \bibinfo{author}{\bibfnamefont{J.}~\bibnamefont{Amato-Grill}},
  \bibinfo{author}{\bibfnamefont{I.}~\bibnamefont{Dimitrova}},
  \bibinfo{author}{\bibfnamefont{W.~W.} \bibnamefont{Ho}},
  \bibinfo{author}{\bibfnamefont{E.}~\bibnamefont{Demler}}, \bibnamefont{and}
  \bibinfo{author}{\bibfnamefont{W.}~\bibnamefont{Ketterle}},
  \bibinfo{journal}{Nature} \textbf{\bibinfo{volume}{588}},
  \bibinfo{pages}{403} (\bibinfo{year}{2020}),
  \urlprefix\url{https://doi.org/10.1038/s41586-020-3033-y}.

\bibitem[{\citenamefont{Chomaz et~al.}(2022)\citenamefont{Chomaz,
  Ferrier-Barbut, Ferlaino, Laburthe-Tolra, Lev, and Pfau}}]{Chomazetal2022}
\bibinfo{author}{\bibfnamefont{L.}~\bibnamefont{Chomaz}},
  \bibinfo{author}{\bibfnamefont{I.}~\bibnamefont{Ferrier-Barbut}},
  \bibinfo{author}{\bibfnamefont{F.}~\bibnamefont{Ferlaino}},
  \bibinfo{author}{\bibfnamefont{B.}~\bibnamefont{Laburthe-Tolra}},
  \bibinfo{author}{\bibfnamefont{B.~L.} \bibnamefont{Lev}}, \bibnamefont{and}
  \bibinfo{author}{\bibfnamefont{T.}~\bibnamefont{Pfau}},
  \emph{\bibinfo{title}{Dipolar physics: A review of experiments with magnetic
  quantum gases}} (\bibinfo{year}{2022}),
  \urlprefix\url{https://arxiv.org/abs/2201.02672}.

\bibitem[{\citenamefont{Christakis et~al.}(2023)\citenamefont{Christakis,
  Rosenberg, Raj, Chi, Morningstar, Huse, Yan, and Bakr}}]{Christakisetal2023}
\bibinfo{author}{\bibfnamefont{L.}~\bibnamefont{Christakis}},
  \bibinfo{author}{\bibfnamefont{J.~S.} \bibnamefont{Rosenberg}},
  \bibinfo{author}{\bibfnamefont{R.}~\bibnamefont{Raj}},
  \bibinfo{author}{\bibfnamefont{S.}~\bibnamefont{Chi}},
  \bibinfo{author}{\bibfnamefont{A.}~\bibnamefont{Morningstar}},
  \bibinfo{author}{\bibfnamefont{D.~A.} \bibnamefont{Huse}},
  \bibinfo{author}{\bibfnamefont{Z.~Z.} \bibnamefont{Yan}}, \bibnamefont{and}
  \bibinfo{author}{\bibfnamefont{W.~S.} \bibnamefont{Bakr}},
  \bibinfo{journal}{Nature} \textbf{\bibinfo{volume}{614}}, \bibinfo{pages}{64}
  (\bibinfo{year}{2023}), ISSN \bibinfo{issn}{1476-4687},
  \urlprefix\url{https://doi.org/10.1038/s41586-022-05558-4}.

\bibitem[{\citenamefont{Monroe et~al.}(2021)\citenamefont{Monroe, Campbell,
  Duan, Gong, Gorshkov, Hess, Islam, Kim, Linke, Pagano
  et~al.}}]{Monroe2021RMP}
\bibinfo{author}{\bibfnamefont{C.}~\bibnamefont{Monroe}},
  \bibinfo{author}{\bibfnamefont{W.~C.} \bibnamefont{Campbell}},
  \bibinfo{author}{\bibfnamefont{L.-M.} \bibnamefont{Duan}},
  \bibinfo{author}{\bibfnamefont{Z.-X.} \bibnamefont{Gong}},
  \bibinfo{author}{\bibfnamefont{A.~V.} \bibnamefont{Gorshkov}},
  \bibinfo{author}{\bibfnamefont{P.~W.} \bibnamefont{Hess}},
  \bibinfo{author}{\bibfnamefont{R.}~\bibnamefont{Islam}},
  \bibinfo{author}{\bibfnamefont{K.}~\bibnamefont{Kim}},
  \bibinfo{author}{\bibfnamefont{N.~M.} \bibnamefont{Linke}},
  \bibinfo{author}{\bibfnamefont{G.}~\bibnamefont{Pagano}},
  \bibnamefont{et~al.}, \bibinfo{journal}{Rev. Mod. Phys.}
  \textbf{\bibinfo{volume}{93}}, \bibinfo{pages}{025001}
  (\bibinfo{year}{2021}),
  \urlprefix\url{https://link.aps.org/doi/10.1103/RevModPhys.93.025001}.

\bibitem[{\citenamefont{Browaeys and Lahaye}(2020)}]{BrowaeysL2020}
\bibinfo{author}{\bibfnamefont{A.}~\bibnamefont{Browaeys}} \bibnamefont{and}
  \bibinfo{author}{\bibfnamefont{T.}~\bibnamefont{Lahaye}},
  \bibinfo{journal}{Nat. Phys.} \textbf{\bibinfo{volume}{16}},
  \bibinfo{pages}{132} (\bibinfo{year}{2020}),
  \urlprefix\url{https://doi.org/10.1038/s41567-019-0733-z}.

\bibitem[{\citenamefont{Garc\'ia-Ripoll}(2022)}]{Juanjobook}
\bibinfo{author}{\bibfnamefont{J.}~\bibnamefont{Garc\'ia-Ripoll}},
  \emph{\bibinfo{title}{{Quantum Information and Quantum Optics with
  Superconducting Circuits}}} (\bibinfo{publisher}{Cambridge University Press},
  \bibinfo{year}{2022}).

\bibitem[{\citenamefont{Levy}(1997)}]{Levybook}
\bibinfo{author}{\bibfnamefont{L.~P.} \bibnamefont{Levy}},
  \emph{\bibinfo{title}{Magnetism and Superconductivity}}
  (\bibinfo{publisher}{Springer}, \bibinfo{year}{1997}).

\bibitem[{\citenamefont{Anderson}(1997)}]{Anderson1997}
\bibinfo{author}{\bibfnamefont{P.~W.} \bibnamefont{Anderson}},
  \emph{\bibinfo{title}{{Basic Notions of Condensed Matter Physics}}}
  (\bibinfo{publisher}{Taylor \& Francis}, \bibinfo{address}{Boca Raton (FL)},
  \bibinfo{year}{1997}).

\bibitem[{\citenamefont{La\"uchli et~al.}(2016)\citenamefont{La\"uchli,
  Schuler, and Wietek}}]{Lauchli2016}
\bibinfo{author}{\bibfnamefont{A.~M.} \bibnamefont{La\"uchli}},
  \bibinfo{author}{\bibfnamefont{M.}~\bibnamefont{Schuler}}, \bibnamefont{and}
  \bibinfo{author}{\bibfnamefont{A.}~\bibnamefont{Wietek}}, in
  \emph{\bibinfo{booktitle}{Quantum Materials: Experiments and Theory -
  Modeling and Simulation, vol. 6}}, edited by
  \bibinfo{editor}{\bibfnamefont{E.}~\bibnamefont{Pavarini}},
  \bibinfo{editor}{\bibfnamefont{E.}~\bibnamefont{Koch}},
  \bibinfo{editor}{\bibfnamefont{J.}~\bibnamefont{van~den Brink}},
  \bibnamefont{and} \bibinfo{editor}{\bibfnamefont{G.}~\bibnamefont{Sawatzky}}
  (\bibinfo{publisher}{Schriften des Forschungszentrums J\"ulich},
  \bibinfo{address}{J\"ulich}, \bibinfo{year}{2016}),
  \urlprefix\url{http://hdl.handle.net/2128/12467}.

\bibitem[{\citenamefont{Tasaki}(2018)}]{Tasaki2018JSP}
\bibinfo{author}{\bibfnamefont{H.}~\bibnamefont{Tasaki}}, \bibinfo{journal}{J.
  Stat. Phys.} \textbf{\bibinfo{volume}{174}}, \bibinfo{pages}{735}
  (\bibinfo{year}{2018}),
  \urlprefix\url{https://doi.org/10.1007/s10955-018-2193-8}.

\bibitem[{\citenamefont{Song et~al.}(2011)\citenamefont{Song, Laflorencie,
  Rachel, and Le~Hur}}]{Songetal2011}
\bibinfo{author}{\bibfnamefont{H.~F.} \bibnamefont{Song}},
  \bibinfo{author}{\bibfnamefont{N.}~\bibnamefont{Laflorencie}},
  \bibinfo{author}{\bibfnamefont{S.}~\bibnamefont{Rachel}}, \bibnamefont{and}
  \bibinfo{author}{\bibfnamefont{K.}~\bibnamefont{Le~Hur}},
  \bibinfo{journal}{Phys. Rev. B} \textbf{\bibinfo{volume}{83}},
  \bibinfo{pages}{224410} (\bibinfo{year}{2011}),
  \urlprefix\url{https://link.aps.org/doi/10.1103/PhysRevB.83.224410}.

\bibitem[{\citenamefont{Fr\'erot and Roscilde}(2015)}]{FrerotR2015}
\bibinfo{author}{\bibfnamefont{I.}~\bibnamefont{Fr\'erot}} \bibnamefont{and}
  \bibinfo{author}{\bibfnamefont{T.}~\bibnamefont{Roscilde}},
  \bibinfo{journal}{Phys. Rev. B} \textbf{\bibinfo{volume}{92}},
  \bibinfo{pages}{115129} (\bibinfo{year}{2015}),
  \urlprefix\url{https://link.aps.org/doi/10.1103/PhysRevB.92.115129}.

\bibitem[{\citenamefont{Fr\'erot et~al.}(2017)\citenamefont{Fr\'erot, Naldesi,
  and Roscilde}}]{Frerot2017PRB}
\bibinfo{author}{\bibfnamefont{I.}~\bibnamefont{Fr\'erot}},
  \bibinfo{author}{\bibfnamefont{P.}~\bibnamefont{Naldesi}}, \bibnamefont{and}
  \bibinfo{author}{\bibfnamefont{T.}~\bibnamefont{Roscilde}},
  \bibinfo{journal}{Phys. Rev. B} \textbf{\bibinfo{volume}{95}},
  \bibinfo{pages}{245111} (\bibinfo{year}{2017}),
  \urlprefix\url{https://link.aps.org/doi/10.1103/PhysRevB.95.245111}.

\bibitem[{\citenamefont{Takahashi}(1989)}]{Takahashi1989}
\bibinfo{author}{\bibfnamefont{M.}~\bibnamefont{Takahashi}},
  \bibinfo{journal}{Phys. Rev. B} \textbf{\bibinfo{volume}{40}},
  \bibinfo{pages}{2494} (\bibinfo{year}{1989}),
  \urlprefix\url{https://link.aps.org/doi/10.1103/PhysRevB.40.2494}.

\bibitem[{\citenamefont{Zhong and Sorella}(1993)}]{ZhongS1993}
\bibinfo{author}{\bibfnamefont{Q.~F.} \bibnamefont{Zhong}} \bibnamefont{and}
  \bibinfo{author}{\bibfnamefont{S.}~\bibnamefont{Sorella}},
  \bibinfo{journal}{Europhysics Letters ({EPL})} \textbf{\bibinfo{volume}{21}},
  \bibinfo{pages}{629} (\bibinfo{year}{1993}),
  \urlprefix\url{https://doi.org/10.1209/0295-5075/21/5/021}.

\bibitem[{\citenamefont{Trumper et~al.}(2000)\citenamefont{Trumper, Capriotti,
  and Sorella}}]{Trumperetal2000}
\bibinfo{author}{\bibfnamefont{A.~E.} \bibnamefont{Trumper}},
  \bibinfo{author}{\bibfnamefont{L.}~\bibnamefont{Capriotti}},
  \bibnamefont{and} \bibinfo{author}{\bibfnamefont{S.}~\bibnamefont{Sorella}},
  \bibinfo{journal}{Phys. Rev. B} \textbf{\bibinfo{volume}{61}},
  \bibinfo{pages}{11529} (\bibinfo{year}{2000}),
  \urlprefix\url{https://link.aps.org/doi/10.1103/PhysRevB.61.11529}.

\bibitem[{\citenamefont{Capriotti}(2003)}]{Capriotti2003}
\bibinfo{author}{\bibfnamefont{L.}~\bibnamefont{Capriotti}},
  \bibinfo{journal}{International Journal of Modern Physics B}
  \textbf{\bibinfo{volume}{17}}, \bibinfo{pages}{4819} (\bibinfo{year}{2003}),
  \eprint{https://doi.org/10.1142/S0217979203023148},
  \urlprefix\url{https://doi.org/10.1142/S0217979203023148}.

\bibitem[{\citenamefont{Roscilde et~al.}(2023)\citenamefont{Roscilde, Comparin,
  and Mezzacapo}}]{Roscildeetal2023}
\bibinfo{author}{\bibfnamefont{T.}~\bibnamefont{Roscilde}},
  \bibinfo{author}{\bibfnamefont{T.}~\bibnamefont{Comparin}}, \bibnamefont{and}
  \bibinfo{author}{\bibfnamefont{F.}~\bibnamefont{Mezzacapo}},
  \emph{\bibinfo{title}{Entangling dynamics from effective rotor/spin-wave
  separation in u(1)-symmetric quantum spin models}} (\bibinfo{year}{2023}),
  \urlprefix\url{https://arxiv.org/abs/2302.09271}.

\bibitem[{\citenamefont{Bruno}(2001)}]{Bruno2001}
\bibinfo{author}{\bibfnamefont{P.}~\bibnamefont{Bruno}},
  \bibinfo{journal}{Phys. Rev. Lett.} \textbf{\bibinfo{volume}{87}},
  \bibinfo{pages}{137203} (\bibinfo{year}{2001}),
  \urlprefix\url{https://link.aps.org/doi/10.1103/PhysRevLett.87.137203}.

\bibitem[{\citenamefont{Chen et~al.}(2022)\citenamefont{Chen, Bornet, Bintz,
  Emperauger, Leclerc, Liu, Scholl, Barredo, Hauschild, Chatterjee
  et~al.}}]{Chenetal2022}
\bibinfo{author}{\bibfnamefont{C.}~\bibnamefont{Chen}},
  \bibinfo{author}{\bibfnamefont{G.}~\bibnamefont{Bornet}},
  \bibinfo{author}{\bibfnamefont{M.}~\bibnamefont{Bintz}},
  \bibinfo{author}{\bibfnamefont{G.}~\bibnamefont{Emperauger}},
  \bibinfo{author}{\bibfnamefont{L.}~\bibnamefont{Leclerc}},
  \bibinfo{author}{\bibfnamefont{V.~S.} \bibnamefont{Liu}},
  \bibinfo{author}{\bibfnamefont{P.}~\bibnamefont{Scholl}},
  \bibinfo{author}{\bibfnamefont{D.}~\bibnamefont{Barredo}},
  \bibinfo{author}{\bibfnamefont{J.}~\bibnamefont{Hauschild}},
  \bibinfo{author}{\bibfnamefont{S.}~\bibnamefont{Chatterjee}},
  \bibnamefont{et~al.}, \emph{\bibinfo{title}{Continuous symmetry breaking in a
  two-dimensional rydberg array}} (\bibinfo{year}{2022}),
  \urlprefix\url{https://arxiv.org/abs/2207.12930}.

\bibitem[{\citenamefont{Holstein and Primakoff}(1940)}]{HP1940}
\bibinfo{author}{\bibfnamefont{T.}~\bibnamefont{Holstein}} \bibnamefont{and}
  \bibinfo{author}{\bibfnamefont{H.}~\bibnamefont{Primakoff}},
  \bibinfo{journal}{Phys. Rev.} \textbf{\bibinfo{volume}{58}},
  \bibinfo{pages}{1098} (\bibinfo{year}{1940}),
  \urlprefix\url{https://link.aps.org/doi/10.1103/PhysRev.58.1098}.

\bibitem[{\citenamefont{Pitaevskii and Stringari}(2016)}]{PitaevskiiStringari}
\bibinfo{author}{\bibfnamefont{L.~P.} \bibnamefont{Pitaevskii}}
  \bibnamefont{and}
  \bibinfo{author}{\bibfnamefont{S.}~\bibnamefont{Stringari}},
  \emph{\bibinfo{title}{Bose-Einstein Condensation}}
  (\bibinfo{publisher}{Oxford}, \bibinfo{year}{2016}).

\bibitem[{\citenamefont{Kitagawa and Ueda}(1993)}]{Kitagawa1993PRA}
\bibinfo{author}{\bibfnamefont{M.}~\bibnamefont{Kitagawa}} \bibnamefont{and}
  \bibinfo{author}{\bibfnamefont{M.}~\bibnamefont{Ueda}},
  \bibinfo{journal}{Phys. Rev. A} \textbf{\bibinfo{volume}{47}},
  \bibinfo{pages}{5138} (\bibinfo{year}{1993}),
  \urlprefix\url{https://link.aps.org/doi/10.1103/PhysRevA.47.5138}.

\bibitem[{\citenamefont{Fr\'esard}(2015)}]{Fresard2015}
\bibinfo{author}{\bibfnamefont{R.}~\bibnamefont{Fr\'esard}}, in
  \emph{\bibinfo{booktitle}{Many-Body Physics: From Kondo to Hubbard - Modeling
  and Simulation, vol. 6}}, edited by
  \bibinfo{editor}{\bibfnamefont{E.}~\bibnamefont{Pavarini}},
  \bibinfo{editor}{\bibfnamefont{E.}~\bibnamefont{Koch}}, \bibnamefont{and}
  \bibinfo{editor}{\bibfnamefont{P.}~\bibnamefont{Coleman}}
  (\bibinfo{publisher}{Schriften des Forschungszentrums J\"ulich},
  \bibinfo{address}{J\"ulich}, \bibinfo{year}{2015}),
  \urlprefix\url{https://juser.fz-juelich.de/record/205123/files/correl15.pdf}.

\bibitem[{\citenamefont{Latorre et~al.}(2005)\citenamefont{Latorre, Or\'us,
  Rico, and Vidal}}]{Latorreetal2005}
\bibinfo{author}{\bibfnamefont{J.~I.} \bibnamefont{Latorre}},
  \bibinfo{author}{\bibfnamefont{R.}~\bibnamefont{Or\'us}},
  \bibinfo{author}{\bibfnamefont{E.}~\bibnamefont{Rico}}, \bibnamefont{and}
  \bibinfo{author}{\bibfnamefont{J.}~\bibnamefont{Vidal}},
  \bibinfo{journal}{Phys. Rev. A} \textbf{\bibinfo{volume}{71}},
  \bibinfo{pages}{064101} (\bibinfo{year}{2005}),
  \urlprefix\url{https://link.aps.org/doi/10.1103/PhysRevA.71.064101}.

\bibitem[{\citenamefont{Sylju\aa{}sen and Sandvik}(2002)}]{Syljuasen2002PRE}
\bibinfo{author}{\bibfnamefont{O.~F.} \bibnamefont{Sylju\aa{}sen}}
  \bibnamefont{and} \bibinfo{author}{\bibfnamefont{A.~W.}
  \bibnamefont{Sandvik}}, \bibinfo{journal}{Phys. Rev. E}
  \textbf{\bibinfo{volume}{66}}, \bibinfo{pages}{046701}
  (\bibinfo{year}{2002}),
  \urlprefix\url{http://link.aps.org/doi/10.1103/PhysRevE.66.046701}.

\bibitem[{\citenamefont{Humeniuk and Roscilde}(2012)}]{HumeniukR2012}
\bibinfo{author}{\bibfnamefont{S.}~\bibnamefont{Humeniuk}} \bibnamefont{and}
  \bibinfo{author}{\bibfnamefont{T.}~\bibnamefont{Roscilde}},
  \bibinfo{journal}{Phys. Rev. B} \textbf{\bibinfo{volume}{86}},
  \bibinfo{pages}{235116} (\bibinfo{year}{2012}),
  \urlprefix\url{http://link.aps.org/doi/10.1103/PhysRevB.86.235116}.

\bibitem[{\citenamefont{Metlitski and Grover}(2011)}]{GroverM2011}
\bibinfo{author}{\bibfnamefont{M.~A.} \bibnamefont{Metlitski}}
  \bibnamefont{and} \bibinfo{author}{\bibfnamefont{T.}~\bibnamefont{Grover}},
  \emph{\bibinfo{title}{Entanglement entropy of systems with spontaneously
  broken continuous symmetry}} (\bibinfo{year}{2011}),
  \urlprefix\url{https://arxiv.org/abs/1112.5166}.

\bibitem[{\citenamefont{Fr\'erot et~al.}(2018)\citenamefont{Fr\'erot, Naldesi,
  and Roscilde}}]{Frerot2018PRL}
\bibinfo{author}{\bibfnamefont{I.}~\bibnamefont{Fr\'erot}},
  \bibinfo{author}{\bibfnamefont{P.}~\bibnamefont{Naldesi}}, \bibnamefont{and}
  \bibinfo{author}{\bibfnamefont{T.}~\bibnamefont{Roscilde}},
  \bibinfo{journal}{Phys. Rev. Lett.} \textbf{\bibinfo{volume}{120}},
  \bibinfo{pages}{050401} (\bibinfo{year}{2018}),
  \urlprefix\url{https://link.aps.org/doi/10.1103/PhysRevLett.120.050401}.

\bibitem[{\citenamefont{Comparin
  et~al.}(2022{\natexlab{a}})\citenamefont{Comparin, Mezzacapo, and
  Roscilde}}]{Comparin2022PRA}
\bibinfo{author}{\bibfnamefont{T.}~\bibnamefont{Comparin}},
  \bibinfo{author}{\bibfnamefont{F.}~\bibnamefont{Mezzacapo}},
  \bibnamefont{and} \bibinfo{author}{\bibfnamefont{T.}~\bibnamefont{Roscilde}},
  \bibinfo{journal}{Phys. Rev. A} \textbf{\bibinfo{volume}{105}},
  \bibinfo{pages}{022625} (\bibinfo{year}{2022}{\natexlab{a}}),
  \urlprefix\url{https://link.aps.org/doi/10.1103/PhysRevA.105.022625}.

\bibitem[{\citenamefont{Comparin
  et~al.}(2022{\natexlab{b}})\citenamefont{Comparin, Mezzacapo, and
  Roscilde}}]{Comparin2022PRL}
\bibinfo{author}{\bibfnamefont{T.}~\bibnamefont{Comparin}},
  \bibinfo{author}{\bibfnamefont{F.}~\bibnamefont{Mezzacapo}},
  \bibnamefont{and} \bibinfo{author}{\bibfnamefont{T.}~\bibnamefont{Roscilde}},
  \bibinfo{journal}{Phys. Rev. Lett.} \textbf{\bibinfo{volume}{129}},
  \bibinfo{pages}{150503} (\bibinfo{year}{2022}{\natexlab{b}}),
  \urlprefix\url{https://link.aps.org/doi/10.1103/PhysRevLett.129.150503}.

\bibitem[{\citenamefont{Wineland et~al.}(1994)\citenamefont{Wineland,
  Bollinger, Itano, and Heinzen}}]{Wineland1994PRA}
\bibinfo{author}{\bibfnamefont{D.~J.} \bibnamefont{Wineland}},
  \bibinfo{author}{\bibfnamefont{J.~J.} \bibnamefont{Bollinger}},
  \bibinfo{author}{\bibfnamefont{W.~M.} \bibnamefont{Itano}}, \bibnamefont{and}
  \bibinfo{author}{\bibfnamefont{D.~J.} \bibnamefont{Heinzen}},
  \bibinfo{journal}{Phys. Rev. A} \textbf{\bibinfo{volume}{50}},
  \bibinfo{pages}{67} (\bibinfo{year}{1994}),
  \urlprefix\url{https://link.aps.org/doi/10.1103/PhysRevA.50.67}.

\bibitem[{\citenamefont{S{\o}rensen et~al.}(2001)\citenamefont{S{\o}rensen,
  Duan, Cirac, and Zoller}}]{Sorensen2001}
\bibinfo{author}{\bibfnamefont{A.}~\bibnamefont{S{\o}rensen}},
  \bibinfo{author}{\bibfnamefont{L.-M.} \bibnamefont{Duan}},
  \bibinfo{author}{\bibfnamefont{J.~I.} \bibnamefont{Cirac}}, \bibnamefont{and}
  \bibinfo{author}{\bibfnamefont{P.}~\bibnamefont{Zoller}},
  \bibinfo{journal}{Nature} \textbf{\bibinfo{volume}{409}}, \bibinfo{pages}{63}
  (\bibinfo{year}{2001}), \urlprefix\url{https://doi.org/10.1038/35051038}.

\bibitem[{\citenamefont{Menu and Roscilde}(2018)}]{MenuR2018}
\bibinfo{author}{\bibfnamefont{R.}~\bibnamefont{Menu}} \bibnamefont{and}
  \bibinfo{author}{\bibfnamefont{T.}~\bibnamefont{Roscilde}},
  \bibinfo{journal}{Phys. Rev. B} \textbf{\bibinfo{volume}{98}},
  \bibinfo{pages}{205145} (\bibinfo{year}{2018}),
  \urlprefix\url{https://link.aps.org/doi/10.1103/PhysRevB.98.205145}.

\bibitem[{\citenamefont{Menu and Roscilde}(2020)}]{MenuR2020}
\bibinfo{author}{\bibfnamefont{R.}~\bibnamefont{Menu}} \bibnamefont{and}
  \bibinfo{author}{\bibfnamefont{T.}~\bibnamefont{Roscilde}},
  \bibinfo{journal}{Phys. Rev. Lett.} \textbf{\bibinfo{volume}{124}},
  \bibinfo{pages}{130604} (\bibinfo{year}{2020}),
  \urlprefix\url{https://link.aps.org/doi/10.1103/PhysRevLett.124.130604}.

\bibitem[{\citenamefont{Cevolani et~al.}(2018)\citenamefont{Cevolani, Despres,
  Carleo, Tagliacozzo, and Sanchez-Palencia}}]{Cevolanietal2019}
\bibinfo{author}{\bibfnamefont{L.}~\bibnamefont{Cevolani}},
  \bibinfo{author}{\bibfnamefont{J.}~\bibnamefont{Despres}},
  \bibinfo{author}{\bibfnamefont{G.}~\bibnamefont{Carleo}},
  \bibinfo{author}{\bibfnamefont{L.}~\bibnamefont{Tagliacozzo}},
  \bibnamefont{and}
  \bibinfo{author}{\bibfnamefont{L.}~\bibnamefont{Sanchez-Palencia}},
  \bibinfo{journal}{Phys. Rev. B} \textbf{\bibinfo{volume}{98}},
  \bibinfo{pages}{024302} (\bibinfo{year}{2018}),
  \urlprefix\url{https://link.aps.org/doi/10.1103/PhysRevB.98.024302}.

\bibitem[{\citenamefont{Lerose et~al.}(2019)\citenamefont{Lerose, \ifmmode
  \check{Z}\else \v{Z}\fi{}unkovi\ifmmode~\check{c}\else \v{c}\fi{}, Marino,
  Gambassi, and Silva}}]{Lerose2019PRB}
\bibinfo{author}{\bibfnamefont{A.}~\bibnamefont{Lerose}},
  \bibinfo{author}{\bibfnamefont{B.}~\bibnamefont{\ifmmode \check{Z}\else
  \v{Z}\fi{}unkovi\ifmmode~\check{c}\else \v{c}\fi{}}},
  \bibinfo{author}{\bibfnamefont{J.}~\bibnamefont{Marino}},
  \bibinfo{author}{\bibfnamefont{A.}~\bibnamefont{Gambassi}}, \bibnamefont{and}
  \bibinfo{author}{\bibfnamefont{A.}~\bibnamefont{Silva}},
  \bibinfo{journal}{Phys. Rev. B} \textbf{\bibinfo{volume}{99}},
  \bibinfo{pages}{045128} (\bibinfo{year}{2019}),
  \urlprefix\url{https://link.aps.org/doi/10.1103/PhysRevB.99.045128}.

\bibitem[{\citenamefont{Lerose and Pappalardi}(2020)}]{Lerose2020PRR}
\bibinfo{author}{\bibfnamefont{A.}~\bibnamefont{Lerose}} \bibnamefont{and}
  \bibinfo{author}{\bibfnamefont{S.}~\bibnamefont{Pappalardi}},
  \bibinfo{journal}{Phys. Rev. Res.} \textbf{\bibinfo{volume}{2}},
  \bibinfo{pages}{012041} (\bibinfo{year}{2020}),
  \urlprefix\url{https://link.aps.org/doi/10.1103/PhysRevResearch.2.012041}.

\bibitem[{\citenamefont{Walls and Milburn}(2008)}]{WallsMilburn}
\bibinfo{author}{\bibfnamefont{D.~F.} \bibnamefont{Walls}} \bibnamefont{and}
  \bibinfo{author}{\bibfnamefont{G.~J.} \bibnamefont{Milburn}},
  \emph{\bibinfo{title}{Quantum Optics}} (\bibinfo{publisher}{Springer},
  \bibinfo{year}{2008}).

\bibitem[{\citenamefont{Garbe et~al.}(2022)\citenamefont{Garbe, Abah,
  Felicetti, and Puebla}}]{Garbeetal2022}
\bibinfo{author}{\bibfnamefont{L.}~\bibnamefont{Garbe}},
  \bibinfo{author}{\bibfnamefont{O.}~\bibnamefont{Abah}},
  \bibinfo{author}{\bibfnamefont{S.}~\bibnamefont{Felicetti}},
  \bibnamefont{and} \bibinfo{author}{\bibfnamefont{R.}~\bibnamefont{Puebla}},
  \bibinfo{journal}{Quantum Science and Technology}
  \textbf{\bibinfo{volume}{7}}, \bibinfo{pages}{035010} (\bibinfo{year}{2022}),
  \urlprefix\url{https://dx.doi.org/10.1088/2058-9565/ac6ca5}.

\bibitem[{\citenamefont{Thibaut et~al.}(2019)\citenamefont{Thibaut, Roscilde,
  and Mezzacapo}}]{Thibaut2019PRB}
\bibinfo{author}{\bibfnamefont{J.}~\bibnamefont{Thibaut}},
  \bibinfo{author}{\bibfnamefont{T.}~\bibnamefont{Roscilde}}, \bibnamefont{and}
  \bibinfo{author}{\bibfnamefont{F.}~\bibnamefont{Mezzacapo}},
  \bibinfo{journal}{Phys. Rev. B} \textbf{\bibinfo{volume}{100}},
  \bibinfo{pages}{155148} (\bibinfo{year}{2019}),
  \urlprefix\url{https://link.aps.org/doi/10.1103/PhysRevB.100.155148}.

\bibitem[{\citenamefont{Schachenmayer et~al.}(2015)\citenamefont{Schachenmayer,
  Pikovski, and Rey}}]{Schachenmayer2015PRX}
\bibinfo{author}{\bibfnamefont{J.}~\bibnamefont{Schachenmayer}},
  \bibinfo{author}{\bibfnamefont{A.}~\bibnamefont{Pikovski}}, \bibnamefont{and}
  \bibinfo{author}{\bibfnamefont{A.~M.} \bibnamefont{Rey}},
  \bibinfo{journal}{Phys. Rev. X} \textbf{\bibinfo{volume}{5}},
  \bibinfo{pages}{011022} (\bibinfo{year}{2015}),
  \urlprefix\url{https://link.aps.org/doi/10.1103/PhysRevX.5.011022}.

\bibitem[{\citenamefont{Carleo et~al.}(2014)\citenamefont{Carleo, Becca,
  Sanchez-Palencia, Sorella, and Fabrizio}}]{Carleo2014PRA}
\bibinfo{author}{\bibfnamefont{G.}~\bibnamefont{Carleo}},
  \bibinfo{author}{\bibfnamefont{F.}~\bibnamefont{Becca}},
  \bibinfo{author}{\bibfnamefont{L.}~\bibnamefont{Sanchez-Palencia}},
  \bibinfo{author}{\bibfnamefont{S.}~\bibnamefont{Sorella}}, \bibnamefont{and}
  \bibinfo{author}{\bibfnamefont{M.}~\bibnamefont{Fabrizio}},
  \bibinfo{journal}{Phys. Rev. A} \textbf{\bibinfo{volume}{89}},
  \bibinfo{pages}{031602} (\bibinfo{year}{2014}),
  \urlprefix\url{https://link.aps.org/doi/10.1103/PhysRevA.89.031602}.

\bibitem[{\citenamefont{Paeckel et~al.}(2019)\citenamefont{Paeckel, K\"ohler,
  Swoboda, Manmana, Schollw\"ock, and Hubig}}]{Paeckeletal2019}
\bibinfo{author}{\bibfnamefont{S.}~\bibnamefont{Paeckel}},
  \bibinfo{author}{\bibfnamefont{T.}~\bibnamefont{K\"ohler}},
  \bibinfo{author}{\bibfnamefont{A.}~\bibnamefont{Swoboda}},
  \bibinfo{author}{\bibfnamefont{S.~R.} \bibnamefont{Manmana}},
  \bibinfo{author}{\bibfnamefont{U.}~\bibnamefont{Schollw\"ock}},
  \bibnamefont{and} \bibinfo{author}{\bibfnamefont{C.}~\bibnamefont{Hubig}},
  \bibinfo{journal}{Annals of Physics} \textbf{\bibinfo{volume}{411}},
  \bibinfo{pages}{167998} (\bibinfo{year}{2019}), ISSN
  \bibinfo{issn}{0003-4916},
  \urlprefix\url{https://www.sciencedirect.com/science/article/pii/S0003491619302532}.

\end{thebibliography}

\end{document}